\documentclass[11pt,a4paper]{article}
\usepackage{jheppub}

\usepackage{bbm,amsmath,graphicx,amssymb}
\usepackage{bm}
\usepackage{verbatim}
\usepackage{slashed}
\usepackage{multirow}

\newcommand{\bea}{\begin{eqnarray}}
\newcommand{\eea}{\end{eqnarray}}
\newcommand{\bean}{\begin{eqnarray*}}
\newcommand{\eean}{\end{eqnarray*}}

\def\Label#1{\label{#1}%
  \smash{\hbox to0pt{\raise1ex\hbox{\tiny[#1]}\hss}}}

\def\stamp{--- {\bf \today} --- {\bf \jobname.tex}}

\def\fs_#1{\mathfrak{s}(#1)}

\def\BE{\begin{equation}}
\def\EE{\end{equation}}
\def\spa#1.#2{\left\langle#1\,#2\right\rangle}
\def\spb#1.#2{\left[#1\,#2\right]}
\def\lor#1.#2{\left(#1\,#2\right)}

\newcommand\fverb{\setbox\fverbbox=\hbox\bgroup\verb}
\newcommand\fverbdo{\egroup\medskip\noindent%
            \fbox{\unhbox\fverbbox}\ }
\newcommand\fverbit{\egroup\item[\fbox{\unhbox\fverbbox}]}
\newbox\fverbbox

\title{On Genera of Curves from High-loop Generalized Unitarity Cuts}

\author[a]{Rijun Huang}
\author[b]{Yang Zhang}
\affiliation{Niels Bohr International Academy and Discovery Center,
The Niels Bohr Institute, Blegdamsvej 17, DK-2100 Copenhagen \O,
Denmark}
\emailAdd{huang@nbi.dk, zhang@nbi.dk}

\abstract{Generalized unitarity cut of a Feynman diagram generates an
algebraic system of polynomial equations. At high-loop levels,
these equations may define a complex curve or a (hyper-)surface with complicated
topology. We study the curve cases, i.e.,  a $4$-dimensional $L$-loop
diagram with $(4L-1)$ cuts. The topology of a complex curve is classified by its
genus. Hence in this paper,  we use computational algebraic geometry to calculate the genera of curves from two and
three-loop unitarity cuts. The
global structure of degenerate on-shell equations under some
specific kinematic configurations is also sketched. The genus
information can also be used to judge if a unitary cut solution could be
rationally parameterized.
}

\keywords{Generalized unitarity cut, Loop amplitudes, Computational algebraic geometry}

\begin{document}
\maketitle
\section{Introduction}

Systematic approach to the multi-loop scattering amplitude study is
now on the road, based on the fruitful progresses of tree and
one-loop amplitude computations in the past a few years. After
struggling with complicated calculations from Feynman diagrams for
decades, the changes started with a new way of computing tree
amplitudes. By complexifying the amplitude through certain momentum
shifting on the complex plane, the amplitude becomes an analytic
function of single complex variable $A(z)$ with simple poles. The
physical amplitude, defined as the amplitude at $z=0$ of the complex
plane, is then calculated by the residues of $A(z)$. The locus
of residues are the phase space points where propagators are
on-shell. Thus, the amplitude can be obtained by considering only
on-shell diagrams with lower-point tree amplitudes as the input via
Britto-Cachazo-Feng-Witten(BCFW) recursion relation
\cite{Britto:2004ap,Britto:2005fq}. This point of view of scattering
amplitude not only provides an efficient calculation method, but
also greatly deepens our understanding of gauge field theories
\cite{ArkaniHamed:2008gz}.

The breakthrough in tree amplitude calculations also inspires
progresses in loop amplitude calculations. Decades ago, the
unitarity cut method \cite{Landau:1959fi} has already been used to
compute one-loop amplitude from tree amplitude
\cite{Bern:1994cg,Bern:1994zx}. After the finding of BCFW
recursion relation, this method is applicable to practical
calculations \cite{Anastasiou:2006jv,Anastasiou:2006gt}. The
difficulty is resolved because of the simple and compact tree
amplitudes produced by the new method, so the focus of unitarity cut
method is switched to the study of unitarity cut information.

Take an one-loop amplitude as example, schematically it can be
expanded on some one-loop integral basis \cite{Britto:2010xq},
\bea A^{{\rm 1-loop}}=\sum_{i\in ~{\rm
basis}}c_iI_i+\mathcal{R}~.~~~\label{1-loop-basis-exp}\eea
The basis $I_i$'s are scalar integrals, i.e., integrals whose
numerators of the integrand are 1. They can be used universally for
any one-loop amplitudes in renormalizable theories. Expansion
coefficients $c_i$ and the remainder $\mathcal{R}$ are rational
functions of external momenta. By cutting both sides of
\ref{1-loop-basis-exp}, i.e., computing the branch-cut
discontinuities across various kinematical channels, we can get
equalities between expressions in both sides, which are products of
tree amplitudes. The coefficients can be determined by
comparing expressions in both sides. 
Now the one-loop amplitude  calculation is simple, because there are
only finite number of integrals in the basis. For example, in
4-dimensional theory, the integral basis $I_i$ can only be box,
triangle, bubble or tadpole topologies, and the numerators are
always constants. The solution space defined by the unitarity cut of
these diagrams is simple: it is just the solution of one quadratic
equation. In this sense, we can easily find a parametrization for the
loop momentum, and work out the expansion coefficients $c_i$
systematically from quadruple, triple, double and single cuts.  
Another way of extracting coefficients $c_i$ is the
generalized unitarity cut method \cite{Britto:2004nc,Britto:2005ha},
which uses the information of cuts and also contour
integration. 

The main concept of loop-amplitude calculations by unitarity-cut
method is to expand the amplitude onto some known integral basis and
then find the coefficients. So first of all, we need a set of
integral basis. The basis should be large enough so that it is
complete for expanding every loop amplitude, yet it should be as
small as possible to simplify the calculation. It is somehow
difficult to find the integral basis for multi-loop higher-point
amplitudes. A conventional way is using Integrate-By-Parts(IBP)
method \cite{IBP} to find relations among different integrals.
Sometimes the IBP calculation can be heavy, so instead we can define
a set of {\it integrand basis} before integration. The denominator
of the integrand is a product of propagators. The numerator is a
polynomial of Lorentz Invariant Scalar Product (ISP).
Ossola-Papadopoulos-Pittau(OPP) decomposition method
\cite{Ossola:2006us} defines the integrand basis and the
corresponding reduction scheme in the one-loop level. This method
has been successfully applied to one-loop amplitude calculations,
where the amplitude is decomposed into some Master Integrals(MIs)
plus the rational terms \cite{Forde:2007mi, Ellis:2007br,
Kilgore:2007qr, Giele:2008ve, Ossola:2008xq, Badger:2008cm}.

It still takes some time to generalize above methods to multi-loop
amplitude calculations beyond the one-loop level  calculation. The
difficulty is obvious: the  basis of multi-loop amplitudes is far
more complicated than those of one-loop amplitudes. Some works have
been done on the integral basis of diagrams such as two-loop
double-box using IBP method \cite{Gluza:2010ws} or generalized
unitarity cut method \cite{Kosower:2011ty,Larsen:2012sx,
CaronHuot:2012ab, Kleiss:2012yv,Johansson:2012zv,Johansson:2012sf}.
By the latter method, the master contours, which produce the
coefficients of double-box integrals in the basis decomposition, are
uniquely defined \cite{CaronHuot:2012ab}. For more complicated
diagrams, the study of integral basis is very complicated, thus we
can consider the integrand basis first. After works in
\cite{Mastrolia:2011pr,Badger:2012dp}, systematic study of integrand
basis begins with the introduction of computational algebraic
geometry method, for example, Gr\"obner basis, to multi-loop amplitude
calculations \cite{Zhang:2012ce,Mastrolia:2012an}. This method can
determine a small enough yet sufficient set of the integrand basis, and
also analyze the solution space of unitarity-cut equations
systematically for two-loop
\cite{Feng:2012bm,Mastrolia:2012wf,Mastrolia:2012du} and also
three-loop amplitudes \cite{Badger:2012dv}.

The main objects we get, after applying the cuts on loop
amplitudes, are unitarity-cut equations. These equations come from
setting corresponding propagators on-shell, i.e., $D_i=\ell_i^2-m_i^2=0$, where $i$'s ranges  from all propagators
being cut, $\ell_i$ is the loop momentum and $m_i$ is the mass of
propagator. In this paper, we consider only massless theories,
so $m_i=0$.

The studies on unitarity-cut equations are crucial for the
application of unitarity methods. For example, the unitarity-cut
equations for $4D$ massless four-point double-box diagram  contain
six irreducible components, and all the six components should be
parameterized to find the constraints for contours
\cite{Kosower:2011ty}. Furthermore,  Simon and Kasper in
\cite{CaronHuot:2012ab} study the two-loop double-box diagram  with
six massive external legs, and find that the unitarity cut equations
define an elliptic curve, i.e. a genus-one curve. This solution space
can be parameterized by elliptic functions, but not rational
functions. The structure of unitarity cuts for all other kinematics
of $4D$ double-box can be easily illustrated via the degeneracy of the
elliptic curve \cite{CaronHuot:2012ab}.

Unitarity cut equations may be simple, if the order of loop is low
and the number of external legs is small. In these cases, one-variable
complex analysis is sufficient for the analysis of unitarity cut
equations. However, for high-loop diagrams or diagrams with many
legs, the analysis would be complicated. Since the cut equations are
all polynomial equations with complex coefficients, the natural tool
to study them is {\it algebraic geometry}.

The generalized unitarity cut of $L$-loop diagrams may define
discrete points, complex curves or (hyper-)surfaces. All of them can
be studied by algebraic geometry methods. In this paper, we consider
the simplest non-trivial cases: the curves. More explicitly, we
consider the cut equations of $L$-loop diagrams with $4L-1$
propagators in $4$-dimensional theory. Since the topology of a
complex curve is completely characterized by its genus, so we study
the genera explicitly for the curves defined by generalized
unitarity cut from one-loop to three-loop level, i.e. $L=1,2,3$. The
same mathematical approach should work for even higher-loop
diagrams. However, the computation would be more involved.

There are many ways to analyze genus and describe the global
structure of multi-loop unitarity cuts, by algebraic geometry. In this
paper, we mainly use the following  methods,
\begin{itemize}
\item The relation between {\it arithmetic genus} and {\it geometric
genus}. In practice, we {\it birationally} project the curve to a
plane curve. For the latter one, the arithmetic genus can be easily
found, and then we use the relation between arithmetic genus and geometric
genus to find the genus of the original curve.
\item Riemann-Hurwitz formula. For a projection of a curve onto
another one, this formula gives a relation between the genera of
the two curves.
\end{itemize}
The genus of a curve is a {\it
birational} invariant, i.e., invariant under rational
re-parametrization. Furthermore, if equations of unitarity cuts
generate a genus-zero curve, then we can find a rational
parametrization for the solution space. However, if the genus is larger
than zero, the solution space cannot be described by rational
parameters no matter how we choose the coordinates. So information
of genus is the first judgement for the difficulty of an unitarity cut
computation.

Given a diagram, we first consider the kinematics with the maximal
number of massive external legs. We call this case the {\it prime
case}. Its unitarity cut sometimes gives high-genus curves. Then we
consider the {\it degenerate cases}, i.e., the same diagram with
fewer external legs or with several massless external legs. Then
usually we get a {\it reducible curve}, which is the union of
several {\it irreducible}  curves. Each curve's genus is lower than
the one of the prime case. For the double-box case in
\cite{CaronHuot:2012ab}, we show again that the global structure of
unitarity cut is given by the prime case, i.e., if we find the
intersection of curves from the degenerate case and "sew" them
together, we get the topology of the prime case. In this sense, the
genus information is the guideline for the global structure of
unitarity cut solutions for all degenerate kinematics.

Throughout this paper, we find that the genus of high-loop
unitarity cuts can be higher than $1$. For example, for a generic
non-planar two-loop crossed-box diagram with $6$ massive momenta,
the on-shell equations of unitarity cuts generate a complex curve
with genus $3$.

The paper is organized as follows. In section 2, we present some
mathematical background, where basic concepts of algebraic
curves, such as genus and singular points, are introduced. In section
3,4 and 5, we study the genera of algebraic curves from one, two and
three-loop diagrams respectively. Detailed analysis is given for
two-loop diagrams, and some typical diagrams of three-loop are also
carefully studied. Discussions and comments are presented in the
conclusion.

\section{Mathematical preliminaries}
In this section, we review the definition and computation of the
{\it genus} of a complex algebraic curve. For detailed mathematical
proofs, please refer to \cite{MR0463157} \cite{MR2372337}.

There are two types of genera for complex curves, {\it arithmetic
genus} and {\it
  geometric genus}. The topology of a complex curve is characterized
by its geometric genus. Roughly speaking, the geometric genus is the
number of handles of a complex curve. (The rigorous definition of
the two genera will be given in this section.) Throughout the paper,
if not specified, ``genus'' means geometric genus.

We calculate the geometric genus of a complex curve directly by
computational algebraic geometry. The strategies are,
\begin{itemize}
\item Calculate the {\it arithmetic genus} of a complex curve. This is
easily done by polynomial computations, which is automated by the
computational algebraic geometry softwares, for example, 'Macaulay2'
\cite{M2} . If the curve is non-singular, then the {\it arithmetic
genus} equals the {\it geometric genus} and we are done. Otherwise,
we project the curve onto to a plane curve and re-calculate the
arithmetic genus. For the plane curve, the geometric genus is simply
related to the arithmetic genus by singular-point counting and the {\it
blow-up} process.
\item Alternatively, project the curve onto a known curve with
smaller genus. This projection needs not to be {\it birational}.  Then
we use Riemann-Hurwitz formula to determine the genus of the first
curve.
\end{itemize}

\subsection{Projective algebraic curves and the arithmetic genus}

Consider a complex curve $C$ in $\mathbb C^n$, defined by polynomial
equations
\begin{equation}
  \label{eq:1}
  f_1(z_1,\ldots z_n)=\ldots = f_k(z_1,\ldots z_n)=0~.~~~
\end{equation}
The polynomials $f_1,\ldots f_k$ generate an ideal $J=\langle f_1
,\ldots f_k\rangle$. In the formal language, $C$ is the {\it zero locus}
of $J$.

To discuss its topology, we need to extend $C$ to a projective curve
$\mathcal C$ in $\mathbb{CP}^n$. Define homogenous coordinates by
$z_i=Z_i/Z_0$, $\mathcal C$ is defined by homogenous polynomials
\begin{equation}
\label{eq:1}
Z_0^{\deg f} \cdot f(\frac{Z_1}{Z_0},\ldots
\frac{Z_n}{Z_0})\equiv F(Z_0,\ldots, Z_n)=0~,~~~\forall f\in J~.~~~
\end{equation}
All such $F$'s generate a homogenous ideal $I$. In practice, to get $I$, we do not need to find the
homogenous form of all polynomials in $J$. Instead, it is sufficient
to consider the homogenous form of polynomials in the Gr\"obner basis of
$J$, in a graded monomial order \cite{MR2290010}.

Furthermore, we assume that $\mathcal C$ is irreducible, i.e.,
cannot be written as the union of two different curves.
Equivalently, it means that $I$ is a prime ideal \cite{MR0463157}.
For the cases of reducible curves, we consider the genus of each
irreducible component. Irreducible components can be found by {\it primary
  decomposition} \cite{MR0463157},
\begin{equation}
  \label{eq:19}
  I=\bigcap_i I_i,\quad i=1,\ldots,k,
\end{equation}
where all $I_i$'s are prime ideals. Primary decomposition technique
has been used in high-loop unitarity cuts \cite{Zhang:2012ce}
\cite{Badger:2012dv} \cite{Feng:2012bm}.

The {\it arithmetic genus} can be defined by the {\it Hilbert
  polynomial} of the quotient ring $S/I$, where $S=\mathbb
C[Z_0,Z_1,\ldots Z_k]$. $S/I$ is a graded S-module,
\begin{equation}
  \label{eq:2}
  S/I=\bigoplus_{i=0}^{\infty} S_i~.~~~
\end{equation}
There exists a unique polynomial $P(x)$ and a positive integer $N$
such that, for $n>N$, $n\in \mathbb Z$,
 \begin{equation}
  \label{eq:3}
  P(n)=\dim_{\mathbb C} S_n~.~~~
\end{equation}
This polynomial is called {\it Hilbert polynomial} (of the
projective curve $\mathcal C$). Then the arithmetic genus of
$\mathcal C$ is defined to be
\begin{equation}
  \label{eq:4}
  g_A(\mathcal C)\equiv -P(0)+1~.~~~
\end{equation}

In practice, the Hilbert polynomial and arithmetic genus can be
easily calculated by Gr\"obner basis method. In particular, for a
{\it plane curve}, i.e. a curve in $\mathbb{CP}^2$, defined by one
homogenous polynomial with degree $d$, the arithmetic genus is
simply \cite{MR0463157}
\begin{equation}
  \label{eq:5}
  g_A= \frac{1}{2} (d-1)(d-2)~.~~~
\end{equation}

\subsection{Singular points and the geometric genus}
The arithmetic genus is not directly related to the topological
properties of a curve, since intuitively it counts not only the number of
handles but also the {\it singular points}. A singular point on
$\mathcal C$ is a point $(a_0, a_1, .... a_n)$ such that the rank of
the Jacobian matrix,
\begin{equation}
  \label{eq:6}
  \bigg|\bigg|\frac{\partial F_i}{\partial Z_j}(a_0, .... a_n)\bigg|\bigg|~,~~~ 1\leq i \leq
  k~~,~~ 0\leq j \leq n~,~~~
\end{equation}
is less than $n-1$ \cite{MR0463157}, where $F_i$, $1\leq i \leq k$ are
the generators of equations for the curve. A singular point on $\mathcal C$ is {\it
  normal} if all tangent lines on the singular point are distinct.

For a {\it smooth} curve, i.e., an irreducible projective curve
without a singular point, we define the {\it geometric genus} to be
its arithmetic genus \cite{MR0463157}. For an irreducible projective
curve $\mathcal C$ with singular points, there exists a smooth
irreducible projective curve $\mathcal{\tilde C}$, which is the {\it
normalization} of $\mathcal C$ \cite{MR0463157},
\begin{equation}
  \label{eq:7}
  \pi: \mathcal{\tilde C} \to C~.~~~
\end{equation}
We define the {\it geometric genus} of $\mathcal C$ to be the
arithmetic genus of $\mathcal{\tilde C}$. This definition is
consistent with the topological definition of the genus. In
particular, for an irreducible {\it plane curve} with normal singular
points only, the geometric genus is related to the arithmetic genus
as
\begin{eqnarray}
  \label{GGGA}
  g_G&=&g_A-\sum_{P\in \text{Sing}(\mathcal C)} \frac{1}{2}
  \mu_P(\mu_P-1) \\
&=&\frac{1}{2} (d-1)(d-2)-\sum_{P\in \text{Sing}(\mathcal C)}
\frac{1}{2}
  \mu_P(\mu_P-1)~,~~~
\end{eqnarray}
where $Sing (\mathcal C)$ is the collection of all singular points
$P$ on $\mathcal C$. $\mu_P$ is the {\it multiplicity} of $P$, i.e.,
the number of tangent lines at $P$. In practice, $Sing (\mathcal C)$
can be found by computational algebraic geometry softwares, like
Macaulay2 \cite{M2}.

We may frequently re-parametrize a curve, and intuitively, the
topological information should be invariant after
re-parametrization. To realize the re-parametrization process
rigorously, we introduce the concept of {\it birational map}:
A {\it rational} map $r$ from the irreducible curves, $C_1$ and $C_2$, is a map from
a non-empty open set of $C_1$ to $C_2$, such that, in terms of the coordinates, all components
of $f$ are rational functions. A {\it birational} map $f$ between two
irreducible curves is a rational map $f$ from $C_1$ to $C_2$ with an
inverse rational map $g$ from $C_2$ to $C_1$, such that $gf$ is the
identity map on an open set of $C_1$ and $fg$ is the
identity map on an open set of $C_2$. The geometric genus is a birational invariant \cite{MR0463157}.

In particular, since $\mathbb{CP}^1$ has the genus zero, a project
curve with $g_G>0$ is not birational to $\mathbb{CP}^1$. It implies
that there is no rational parametrization for such a curve.

We present some simple examples on the genera of algebraic curves:
\begin{enumerate}
\item $y^2=x^3+1$. This is a cubic plane curve. The corresponding
  projective curve is $Y^2 Z=X^3+Z^3$. There is no singular point on
  the projective curve. The arithmetic genus is $(3-2)(3-1)/2=1$ and
  the geometric genus is also $1$. So this is an elliptic curve.

\item $y^2=x^3+x^2$. This is also a cubic plane curve. The
  corresponding projective curve is $Y^2 Z=X^3+X^2 Z$. There is a
  singular point at $X=0, Y=0, Z=1$ and there are two distinct tangent
  lines at this point: $x-y=0$ and $x+y=0$. So this is an ordinary
  singular point with the multiplicity $2$. The arithmetic genus is
  still $1$ but the geometric genus is $1-2(2-1)/2=0$. So this is not
  an elliptic curve. We can blow up the singular point at
  $(x,y)=(0,0)$ as follows \cite{MR0463157}: define a new variables
  $t$ via $y=x
  t$, and the equation becomes $-x^2 (1-t^2+x)=0$. Factorize it and
  keep the curve part, we get a new curve $1-t^2+x=0$. This new curve
  is birational to the original curve, and is smooth. Since the new
  curve is conic, so we show that the original curve is {\it birationally equivalent} to a
  conic and can be rational parameterized.

\item The projective curve,
  \begin{equation}
    \label{eq:9}
X T-Y Z=0,\quad X^2 Z+Y^3+ Y Z T=0,\quad X Z^2+Y^2 T+Z T^2=0
  \end{equation}
is non-singular \cite{MR2372337} . From Macaulay2, we find that the
arithmetic genus is $2$, so its geometric genus is also $2$.
\end{enumerate}

There are many different ways to calculate the genus of an algebraic
curve. In this paper, for the algebraic curves from unitarity cuts
in two-loop and three-loop orders, we mainly use the relation
between arithmetic genus and geometric genus for the computation.
Explicitly, if the curve is non-singular, then we simply calculate
its arithmetic genus and by the definition, it is the geometric
genus. Otherwise, we birationally project the curve onto a plane,
and then use (\ref{GGGA}) to determine the genus if all singular
points are normal. If some singular points are not normal, we use
the {\it blow up} process to resolve them, as described in
\cite{MR2372337}.

Anther useful formula for curve genus computation is {\it
  Riemann-Hurwitz formula}  \cite{MR0463157}.  Let $f: C_1 \to C_2$ be a covering map between two irreducible complex curves $C_1$
and $C_2$, and $n$ is the degree of $f$. Then $g(C_1)$ and $g(C_2)$,
the genera of $C_1$ and $C_2$ are related by
\begin{equation}
  \label{Riemann-Hurwitz}
  2 g(C_1)-2=n\cdot (2 g(C_2)-2))+\sum_{P\in C_1}(e_P-1)~,~~~
\end{equation}
where $e_P$ is the ramification index of $P$. For all but finite
points on $C_1$, the ramification index equals one so it is just a
finite sum. Therefore if $g(C_2)$ is known, we can get $g(C_1)$ by
counting the ramified points on $C_1$.

\section{Genera of algebraic curves from one-loop diagrams}

For one-loop diagrams, there is only one loop momentum, and it has
$4$ components. Diagrams with $3$ propagators after maximal
unitarity cut define algebraic curves. The only diagram we
should consider is the triangle diagram. The propagators are given
by
\bea
D_0=\ell^2~,~~~D_1=(\ell-p_1)^2~,~~~D_2=(\ell-p_1-p_2)^2~,~~~\eea
where external momenta at each vertices are $p_1,p_2,p_3$, and we
assume they are all massive. We can get $2$ polynomials
$L_1=D_0-D_1$, $L_2=D_0-D_2$, which are linear in $\ell$. If we use
$4$ variables $(x_1,x_2,x_3,x_4)$ to expand the loop momentum, then
$L_1=L_2=0$ are $2$ linear equations in them. By solving these
equations and substituting solutions back to $D_0$, we get a
quadratic polynomials of $2$ variables. The resulting equation
$D_0=0$ is then a conic section, and topologically equivalent to a
genus-0 Riemann sphere.

If some of the external momenta are massless, the on-shell equations
become degenerate. The topological picture can be given by combing another Riemann sphere from the original one, and the result is that
two Riemann spheres are connected at a single point. This agrees with
the analysis of intersection pattern in \cite{Feng:2012bm}.

\section{Genera of algebraic curves from two-loop diagrams}

In this section, we calculate the genera of all curves from the $4D$
two-loop unitarity cuts, for massless theories

For 4-dimensional two-loop diagrams, there are $8$ components from
two loop momenta.
The $8$ variables
associated to the $8$ components of loop momenta could be
defined as the expansion coefficients of loop momenta on a chosen
momentum basis. They can also be defined by Lorentz invariant scalar
product of loop momentum and external momentum. The resulting
propagators, which expressed as polynomials of these $8$ variables,
will be different according to the definitions of momentum basis and
variables. But in general,  they are always quadratic polynomials and only
differ by a linear transformation.

If we apply the maximal unitarity cut on the diagram, all
propagators become on-shell, then we get a polynomial equation
system. These equations define an {\it algebraic set}, i.e. the
solution set of cut equations in the space spanned by $8$ variables.
When considering diagrams with $7$ cuts, the solution set is
described by $7$ equations with respect to $8$ variables. It is an
algebraic curve.

It is not easy to analyze this curve directly from its
defining polynomial equation system. So the first step
is to eliminate as many variables as possible from
the linear equations which can be constructed from $7$ on-shell
equations. In practice, it is always possible to obtain $4$
linear equations. By solving them, we can express $4$ variables
as linear combinations of the remaining $4$ variables. The resulting
$3$ quadratic polynomial equations of $4$ variables is isomorphic to
the original algebraic set. Furthermore,  by coordinate
transformations, we  can rewrite the $3$ equations as a single
meromorphic function of $2$ variables. It defines a plane curve,
which is birationally equivalent to the original algebraic set.
Though the explicit definitions of curves in above three descriptions
are different, the topological invariant objects remain the same. So
we can get a unique result for the genus.

\subsection{Two-loop planar pentagon-triangle diagram, genus 0}

\begin{figure}
\center
  \includegraphics[width=6.2in]{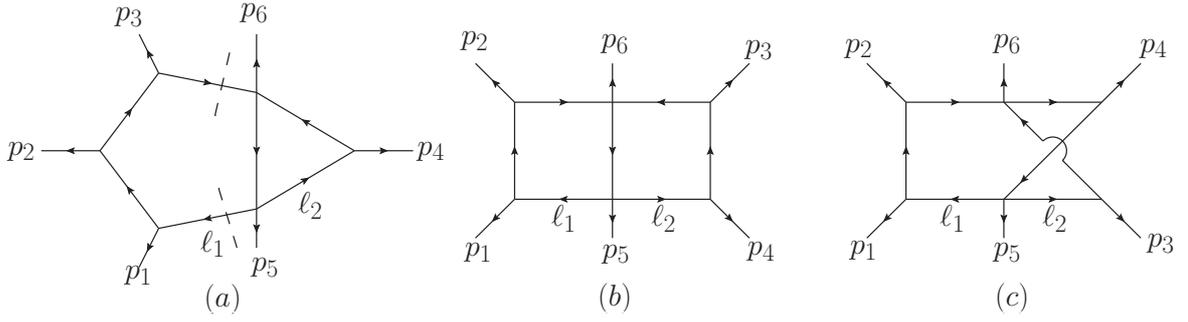}\\
  \caption{Two-loop diagrams with $7$ propagators: (a) planar pentagon-triangle diagram, (b) planar double-box diagram,
  (c) non-planar crossed-box diagram. All external momenta are out-going and massive. The loop momenta are denoted
  by $\ell_1,\ell_2$. }\label{2loop}
\end{figure}

As a warm-up, we first analyze this simple diagram, in
Figure.(\ref{2loop}.a). The $7$
propagators are given by
\bea
&&D_0=\ell_1^2~,~~~D_1=(\ell_1-p_1)^2~,~~~D_2=(\ell_1-p_1-p_2)^2~,~~~D_3=(\ell_1-p_1-p_2-p_3)^2~,~~~\nonumber\\
&&\widetilde{D}_0=\ell_2^2~,~~~\widetilde{D}_1=(\ell_2-p_4)^2~,~~~\widehat{D}_0=(\ell_1+\ell_2+p_5)^2~.~~~\eea
There are $4$ propagators containing $\ell_1$ only, and from the
on-shell equations $D_i=0$, we can completely fix the variables in
$\ell_1$. These equations $D_i=0, i=1,2,3,4$ have two solutions. For
each solution, since $\ell_1$ has already been completely fixed, we
can treat it as a four-vector similar to an external momentum. Thus the
on-shell propagators highlighted by dashed lines in Figure
(\ref{2loop}.a)
can be effectively treated as massless external
momenta. Then for each solution, the result of two-loop planar
pentagon-triangle diagram is the same as the result of the one-loop
triangle diagram. It is associated with a genus-0 Riemann sphere. In some specific kinematic configurations where degeneracy exists,
the topological picture becomes that of two Riemann spheres connected at a
single point.  

\subsection{Two-loop planar double-box diagram, genus up to 1}

It is already known in \cite{CaronHuot:2012ab} that on-shell
equations of the two-loop double-box diagram, with six massive legs,
define an one-dimensional elliptic curve which is associated with a
genus-$1$ torus.

We can reproduce this result by directly computing the arithmetic genus and
singular points of the curve. Taking the convention shown in Figure
(\ref{2loop}.b), the $7$ propagators can be written as,
\bea
&&D_0=\ell_1^2~,~~~D_1=(\ell_1-p_1)^2~,~~~D_2=(\ell_1-p_1-p_2)^2~,~~~\nonumber\\
&&\widetilde{D}_0=\ell_2^2~,~~~\widetilde{D}_1=(\ell_2-p_4)^2~,~~~\widetilde{D}_2=(\ell_2-p_3-p_4)^2~,~~~\widehat{D}_0=(\ell_1+\ell_2+p_5)^2~.~~~\eea

In order to compute the arithmetic genus and singular points, we
first rewrite above equations by expanding all momenta on
a momentum basis. A simple analysis is as follows. We have variables
$(x_1,x_2,x_3,x_4)$ for 4 components of $\ell_1$ and
$(y_1,y_2,y_3,y_4)$ for 4 components of $\ell_2$. Using on-shell
equations $D_i=0, \widetilde{D}_i=0$, we can eliminate $2$ variables
of $x_i$ and $2$ of $y_i$, and get $3$ quadratic polynomials
$Q_1=Q_1(x_1,x_2)$, $Q_2=Q_2(y_1,y_2)$, $Q_3=Q_3(x_1,x_2,y_1,y_2)$.
This can be done by Yang's BasisDet package\cite{Zhang:2012ce}. We can further
eliminate $2$ variables and $2$ equations via Gr\"obner basis method
and get a plane curve. This plane curve has degree $8$, so the
arithmetic genus is
\bea g_A={(d-1)(d-2)\over 2}=21~.~~~\eea
There are $10$ singular points, of which $8$ have the multiplicity
$\mu_{P}=2$ and $2$ have the multiplicity $\mu_{P}=4$, so the geometric
genus is
\bea g_G=g_A-\sum_{P\in Sing(C)}{1\over 2}\mu_P(\mu_P-1)=21-8\times
1-2\times 6=1~.~~~\eea
This result is consistent with that in \cite{CaronHuot:2012ab}.

Another explicit way of getting an equivalent plane curve can be
taken as follows. Through coordinate transformation it is easy to
rewrite $Q_1\to Q'_1=x'_1 x'_2-c_1$, $Q_2\to Q'_2=y'_1y'_2-c_2$, so
we can do the following substitution $x'_1=c_1/x'_2, y'_1=c_2/y'_2$
in $Q_3$. The resulting $Q'_3=n(x'_2,y'_2)/d(x'_2,y'_2)$ is a
meromorphic function which is equivalent to original equations, and the
numerator $n(x'_2,y'_2)$ defines a plane
curve. This plane curve $n(x'_2,y'_2)$ has the degree $4$, so we have
$g_A=3$. There are $2$ singular points of the multiplicity $\mu_P=2$,
thus we get the geometric genus $g_G=3-2=1$. This result again agrees
with that given in \cite{CaronHuot:2012ab}.

\begin{figure}
\center
  \includegraphics[width=6in]{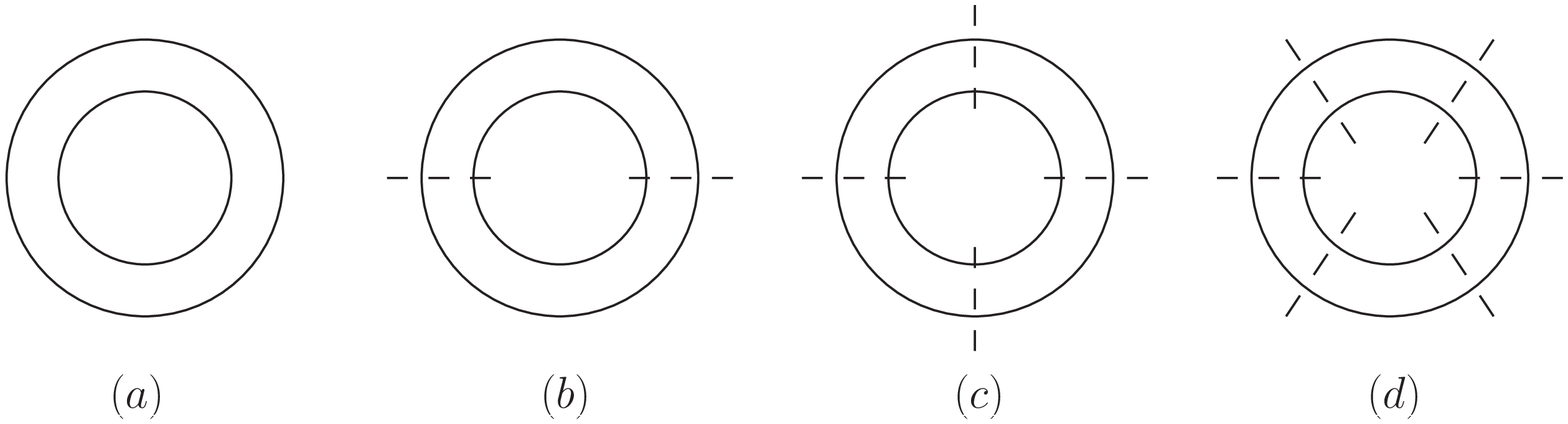}\\
  \caption{Topological pictures of on-shell equations from the two-loop double-box diagram under
specific kinematic configurations. The pictures should be understood
as complex curves, or two-dimensional real surfaces. (a) the curve
is irreducible and the solution set is a torus (b) the curve has 2 irreducible branches,
(c) the curve has 4 irreducible branches, (d)   the curve has 6 irreducible branches. For general kinematics the
curve is genus 1. In degenerate limit, tubes shrink to points along dashed lines. The
resulting Riemann surfaces for each branch can only be a sphere. This
explains why we get Riemann spheres connected by points and linked
in a chain in degenerate limits. }\label{2loopdbox}
\end{figure}

The topological pictures of genus of degenerate algebraic curves
under specific kinematic configurations can be easily found, because
in degenerate limit the tube shrinks to point, and there is only one
handle for the torus. The contraction will break the handle, and we
should get genus-0 Riemann spheres connected at points. Equations of
each irreducible branch can be obtained by Macaulay2 \cite{M2} via
primary decomposition. By computing the arithmetic genus and
singular points, we found that their geometric genera are indeed
zero. The $3$ kinds of contraction shown in Figure (\ref{2loopdbox})
describe topological pictures of degenerate algebraic curves under
all possible specific kinematic configurations. There are $2$, $4$
or $6$ Riemann spheres connected by points along dashed lines where
tubes shrink, and linked adjacently into a chain. This is exactly
the picture described in \cite{CaronHuot:2012ab}.

\subsection{Two-loop non-planar crossed-box diagram, genus up to 3}

For the two-loop non-planar crossed-box diagram, the explicit form of
propagators for generic kinematics is very complicated. Abstractly,
we write the $7$ propagators as,
\bea
&&D_0=\ell_1^2~,~~~D_1=(\ell_1-p_1)^2~,~~~D_2=(\ell_1-p_1-p_2)^2~,~~~\widetilde{D}_0=\ell_2^2~,~~~\nonumber\\
&&\widetilde{D}_1=(\ell_2-p_3)^2~,~~~\widehat{D}_0=(\ell_1+\ell_2+p_5)^2~,~~~\widehat{D}_1=(\ell_1+\ell_2+p_4+p_5)^2~.~~~\eea
From above propagators, we can at most get $4$ polynomials which are
linear in $\ell$,
\bea &&L_1=D_0-D_1=2\ell_1p_1-p_1^2~,~~~L_2=D_0-D_2=2\ell_1(p_1+p_2)-(p_1+p_2)^2~,~~~\nonumber\\
&&L_3=\widetilde{D}_0-\widetilde{D}_1=2\ell_2p_3-p_3^2~,~~~L_4=\widehat{D}_0-\widehat{D}_1=-2(\ell_1+\ell_2+p_5)p_4-p_4^2~,~~~\nonumber\eea
and $3$ polynomials which are quadratic in $\ell$,
\bea
&&Q_1=D_0=\ell_1^2~,~~~Q_2=\widetilde{D}_0=\ell_2^2~,~~~Q_3=\widehat{D}_0-D_0-\widetilde{D}_0=2\ell_1\ell_2+2(\ell_1+\ell_2)p_5+p_5^2~.~~~\nonumber\eea
If we use momentum basis $e_1,e_2,e_3,e_4$ to expand loop momenta as,
\bea \ell_1=x_2 e_1+x_1 e_2+x_4 e_3+x_3 e_4~,~~~\ell_2=y_2 e_1+y_1
e_2+y_4 e_3+y_3 e_4~,~~~\eea
and the $8$ variables are taken as the expansion coefficients
$(x_1,x_2,x_3,x_4)$ and $(y_1,y_2,y_3,y_4)$, then there will be
quadratic terms from products of two loop momenta $\ell_i\ell_j$,
and linear terms from products like $\ell_ip_j$. 
%
%
Systematically we can generate the Gr\"obner basis by $7$
on-shell equations $D_i=\widetilde{D}_i=\widehat{D}_i=0$, and find
all linear terms in the Gr\"obner basis. We can remove as many
variables as possible by solving these linear terms, and express
other terms with remaining variables. In this case, we get an
equivalent but simplified polynomial equation system. A more explicit
way is to solve linear equations $L_i=0$, and express $4$ variables
as linear functions of remaining $4$ variables. Substituting
solutions back to quadratic polynomial equations $Q_i=0$, we get an
alternate algebraic set. It is possible to analyze the
three quadratic equations by algebraic geometry program such as
Macaulay2 \cite{M2}.

Let us consider the two-loop non-planar crossed-box diagram drawn
in Figure (\ref{2loop}.c). All external momenta are general, and we
construct momentum basis $e_i$ from $p_1,p_3$\footnote{Momentum
basis $e_i$ have the following orthogonal and normalization
relations $e_1e_2=e_3e_4=1$, otherwise $0$. The explicit form can be
found by Gram-Schmidt process.}. External momenta are
expanded as
\bea
&&p_1=\alpha_{1}e_1+\alpha_{12}e_2~,~p_1+p_2=\sum_{i=1}^4\alpha_{2i}e_i~,~p_3=\beta_{11}e_1+\beta_{12}e_2~,~~~\nonumber\\
&&p_5=\sum_{i=1}^{4}\gamma_{1i}e_i~,~p_4+p_5=\sum_{i=1}^4\gamma_{2i}e_i~,~p_6=-(p_1+p_2+p_3+p_4+p_5)~,~~~\eea
where $\alpha,\beta,\gamma$ are the projection coefficients of momenta
on the
corresponding momentum basis, for example $\alpha_{11}=p_1e_2,
\alpha_{12}=p_1e_1$. Using this expansion, the four linear
polynomials $L_i$ become
\bea
&&L_1=2(\alpha_{11}x_1+\alpha_{12}x_2-\alpha_{11}\alpha_{12})~,~~~\nonumber\\
&&L_2=2(\alpha_{21}x_1+\alpha_{22}x_2+\alpha_{23}x_3+\alpha_{24}x_4-\alpha_{21}\alpha_{22}-\alpha_{23}\alpha_{24})~,~~~\nonumber\\
&&L_3=2(\beta_{11}y_1+\beta_{12}y_2-\beta_{11}\beta_{12})~,~~~\nonumber\\
&&L_4=-2\sum_{i=1}^4(x_i+y_i)(\gamma_{2i}-\gamma_{1i})-2(\gamma_{21}\gamma_{22}+\gamma_{23}\gamma_{24}-\gamma_{11}\gamma_{12}-\gamma_{13}\gamma_{14})~,~~~\eea
and the three quadratic polynomials $Q_i$ become
\bea &&Q_1=2(x_1x_2+x_3x_4)~,~Q_2=2(y_1y_2+y_3y_4)~,~~~\nonumber\\
&&Q_3=2(x_1y_2+x_2y_1+x_3y_4+x_4y_3)+2\sum_{i=1}^4(x_i+y_i)\gamma_{1i}+2(\gamma_{11}\gamma_{12}+\gamma_{13}\gamma_{14})~.~~~\eea
The equations until now  are still simple, but after solving $L_i=0$
and substituting solutions back to $Q_i$, they will become quadratic
polynomials with complicate coefficients of
$\alpha,\beta,\gamma$. Anyway,
$Q_i$'s are still $3$ quadratic polynomials in $4$ variables, and
they define a one-dimensional curve.

As we have mentioned, different definitions of momentum basis and
variables, and the choices of choosing remaining variables after solving
linear equations, will describe different curves. But they are all
birationally equivalent. Birational invariants will stay
the same in each description. 
%
%
So we project the three
quadratic equations onto a plane curve. By a coordinate transformation, we
can eliminate $2$ equations as well as $2$ variables, and the
resulting equation defines a plane curve describing
an equivalent algebraic set. 
This equation has the degree $d=8$, so the
arithmetic genus is,
\bea g_A={1\over 2}(d-1)(d-2)=21~.~~~\eea
The singular points can be obtained directly from the definition,
i.e., for projective plane curve defined by homogeneous polynomial
$P(x,y,z)$ of degree $d$, the singular points are solutions of
equations
\bea P(x,y,z)=P'_x(x,y,z)=P'_y(x,y,z)=P'_z(x,y,z)=0~.~~~\eea
For this plane curve, there are 18 normal singular points of
the multiplicity $\mu_P=2$, so the geometric genus is again
\bea g_G=g_A-\sum_{P\in Sing(C)}{1\over
2}\mu_P(\mu_P-1)=21-18=3~.~~~\eea

Another way of getting a plane curve from original on-shell equations
is to use the elimination process via Gr\"obner basis method. We can
use BasisDet package to generate on-shell equations, where the
variables are defined as independent Lorentz Invariant Scalar
Products (ISPs) of loop momentum and external momentum. Using these cut
equations and Gr\"obner basis associated with them, we get a plane
curve. The degree of
the plane curve is $8$, so we have $g_A=21$. There are $18$ normal
singular points of the multiplicity $2$, so again we get the
geometric genus $g_G=21-18=3$.

To summarize, the maximal unitarity cut of two-loop non-planar
crossed-box diagram can fix $7$ components of loop momenta, and the
solution set of on-shell equations is described by a free parameter
corresponding to the remaining one degree of freedom. The variety is
associated with a genus-$3$ Riemann surface.

\subsubsection{Degeneracy under specific kinematics}

In the prime case, external momenta $p_i$ are massive and the algebraic set
defined by on-shell equations is irreducible. Under some
specific kinematic configurations, the variety will be decomposed to
many irreducible ideals after primary decomposition. For example, if
$p_1$ is massless in Figure (\ref{2loop}.c), the algebraic set
decomposes into $2$ irreducible ideals, i.e., it has two branches.
Each irreducible branch is associated with a Riemann surface, and it
should have connection with the original genus-$3$ Riemann surface.

\begin{figure}
\center
  \includegraphics[width=6in]{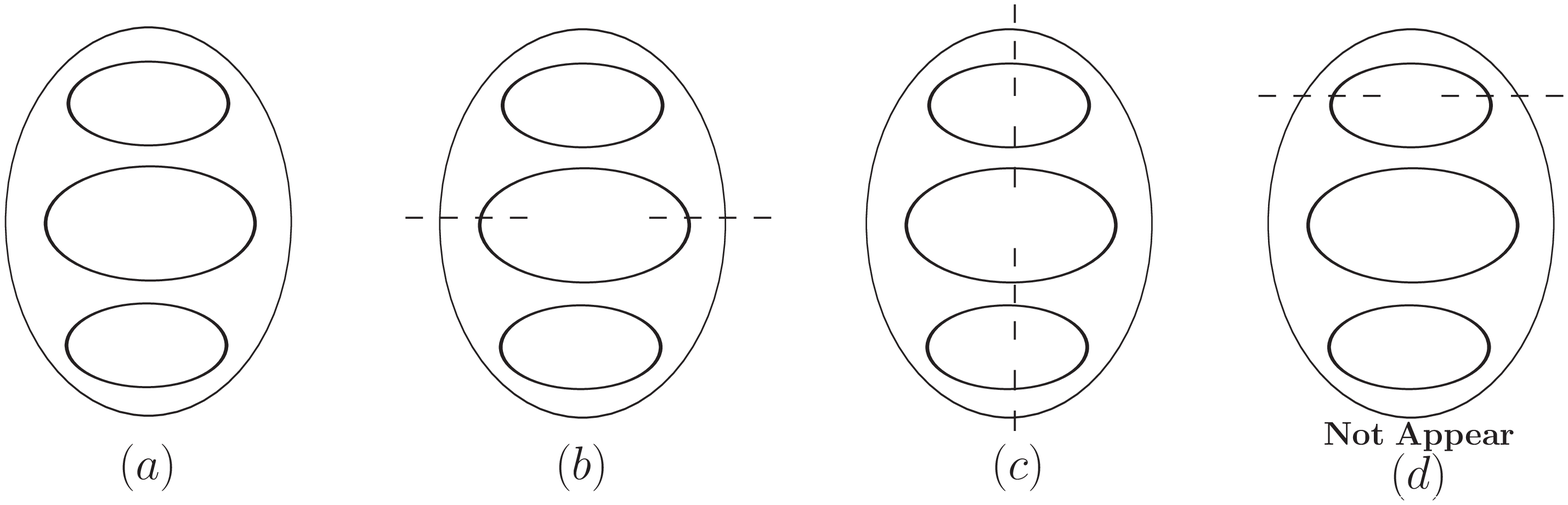}\\
  \caption{Topological pictures of degenerate on-shell equations from non-planar two-loop
  crossing-box diagram under
  specific kinematic configurations. (a) the curve is irreducible, the geometric genus is $3$,
  (b) the curve has 2 irreducible branches, each branch is genus 1, and
  they are connected by two points,
  (c) the curve has 2 irreducible branches, each branch is genus 0, and they are connected by four points,
  (d) no kinematic configuration corresponds to this topological
  picture. 
}
\label{2loopxbox}
\end{figure}

We can compute the geometric genus for each irreducible branch using the
same method in previous section. Let us start from the kinematic
configuration where at least one momentum of $p_1,p_2$ is massless.
Simple calculation shows that for $p_1^2=0$, we have $Q_1=x_3x_4$,
so $Q_1=0$ implies that either $x_3=0$ or $x_4=0$. For $p_2^2=0$, we
have $Q_1=f_1(x_3,x_4)f_2(x_3,x_4)$, where $f_1,f_2$ are linear
polynomials of $x_3,x_4$, and $Q_1=0$ implies that either $f_1=0$ or
$f_2=0$. Combined with $Q_2=0,Q_3=0$, there are two branches. For each branch, we can transfer the 3 equations to
a plane curve by coordinate transformation. This plane curve has the
degree $4$, so the arithmetic genus is $g_A=3$. There are $2$ normal
singular points of the multiplicity $\mu_P=2$. Finally the geometric
genus is $g_G=3-2=1$ for each branch, so they are associated with a
genus-$1$ torus. The two tori intersect at $2$
points \cite{Feng:2012bm}. From these it is easy to obtain that, the
topological picture under this kinematic configuration is given by
contracting two tubes to points along dashed lines, as
shown in Figure (\ref{2loopxbox}.b).
The up and down parts become tori, and they are
connected by $2$ points, which are originally $2$ tubes.

If we consider kinematic configurations where only one of
$p_3,p_4$ is massless, or only one of $p_5,p_6$ is absent
(which means that only one vertex of those attached to
$p_3,p_4,p_5,p_6$ is a three-vertex with chirality), the algebraic set is
reducible, and it has two irreducible branches. For $p_3^2=0$, we
have $Q_2=y_3y_4$, and we can get two branches directly from
$Q_2=0$. For other kinematic configurations, it is not easy to see
the degeneracy directly from on-shell equations, since none of
$Q_1,Q_2,Q_3$ can be obviously factorized into two parts. Then we
generate an ideal from $Q_1,Q_2,Q_3$, get two irreducible
ideals using primary decomposition method of Macaulay2 \cite{M2}. For each
irreducible ideal, we  can obtain a plane curve after coordinate
transformations. Each plane curve has the degree $4$, so the arithmetic
genus is $g_A=3$. There are $3$ normal singular points of
the multiplicity $\mu_P=2$, so we have $g_G=3-3=0$. It means that for
all kinematic configurations considered in this paragraph, each
irreducible branch is associated with a genus-0 Riemann sphere.
Another way of analysis is to get a plane curve directly from
$Q_1=Q_2=Q_3=0$, and the resulting equation has the degree $8$. This
equation can be factorized to $2$ polynomials of the degree $4$, and each
factor is equivalent to one irreducible branch. We can again get
geometric genus $g_G=3-3=0$. These two Riemann spheres intersect at
4 points\cite{Feng:2012bm}, so the possible topological
picture can be given by contracting $4$ tubes to points along dashed lines as shown in Figure (\ref{2loopxbox}.c). The left and
right parts become Riemann spheres, and they
intersect at $4$ points.

Above discussion involves all kinematic configurations where the
algebraic set has two irreducible branches. From a genus-$3$ Riemann
surface, there is another possibility of contracting two tubes to
get two Riemann surfaces as shown in Figure (\ref{2loopxbox}.d), and
the resulting topological picture is: a genus-0 Riemann sphere and
genus-$2$ Riemann surface intersect at two single points. But no
kinematic configuration is found to guarantee the on-shell equations
having such degeneracy. Intuitively, they can be understood as the
consequence of symmetry between two branches. Usually, the two
branches are related by the parity symmetry and they would have the
same genus.


If we denote kinematic condition as,
\begin{itemize}
\item  $K_1$: at least one of $p_1,p_2$ is massless,
\item $K_2^a$: $p_3$ is massless,
\item $K_2^b$: $p_4$ is massless,
\item $K_3^a$: $p_5$ is absent,
\item  $K_3^b$: $p_6$ is absent,
\end{itemize}
then the general rules for degenerate topological picture
can be given as follows: condition $K_1$ will contract the tubes
at dashed lines in Figure (\ref{2loopxbox}.b), and condition
$K_2^a$, $K_2^b$, $K_3^a$ or $K_3^b$ will contract the tubes at
dashed lines in Figure (\ref{2loopxbox}.c). Topological picture of cases
with more than two irreducible branches can be determined by
combining these conditions.

\begin{figure}
\center
  \includegraphics[width=6in]{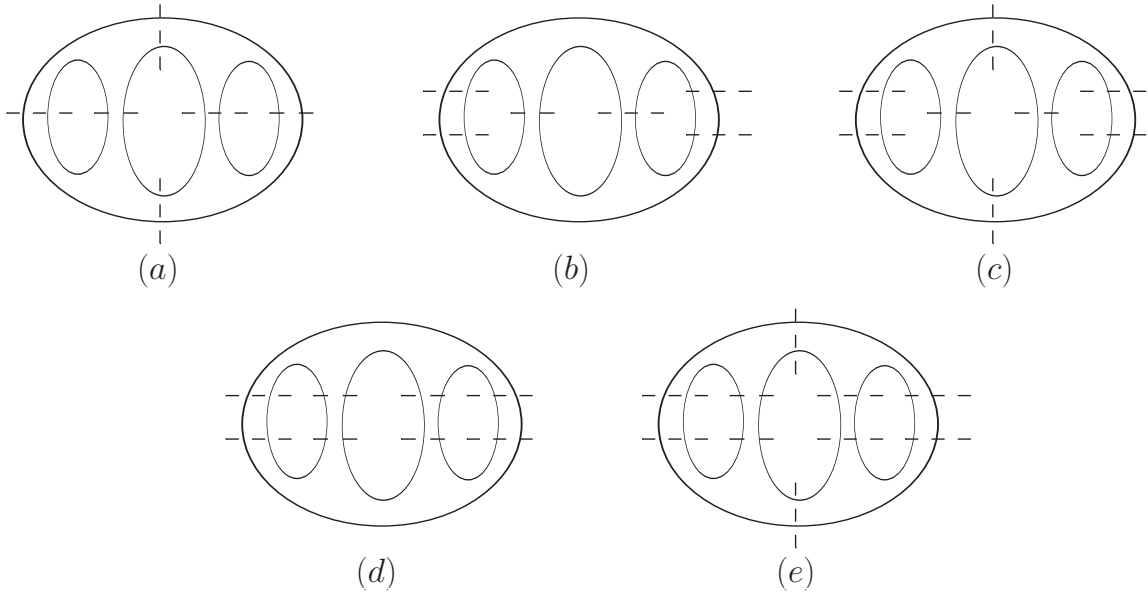}\\
  \caption{Topological pictures of degenerate on-shell equations from non-planar two-loop
  crossing-box diagram under
  specific kinematic configurations, where the solution set has more than $2$ branches. These $5$
  pictures includes the degeneracies under all possible kinematic configurations. Each irreducible
  branch is genus 0 sphere, and they are connected by points along dashed lines where
  tubes have been contracted.}\label{2loopxboxmore}
\end{figure}

Explicitly, if we combine one condition
$K_1$ with one of $K_2^a$, $K_2^b$, $K_3^a$, $K_3^b$, the
topological picture is given in Figure (\ref{2loopxboxmore}.a),
which is the overlap of (\ref{2loopxbox}.b) and (\ref{2loopxbox}.c).
This gives $4$ Riemann spheres, which means that there are $4$
irreducible branches.  The on-shell equations of this kinematic
configuration indeed have $4$ irreducible branches,  and by the
explicit computation, each one is a genus-0 curve,
which agrees with the topological picture.

If we combine one
condition of $K_2^a,K_2^b$ with one of $K_3^a,K_3^b$, the
topological picture is given in Figure (\ref{2loopxboxmore}.b). This
gives $4$ Riemann spheres. It is not hard to find that Figure
(\ref{2loopxboxmore}.c) is given by further combining $K_1$ with
above configuration, and the resulting picture gives $6$ Riemann
spheres. For kinematic configurations with both $K_2^a,K_2^b$ or
both $K_3^a,K_3^b$, the picture is given by overlapping the double copy
of Figure (\ref{2loopxbox}.c), as shown in Figure
(\ref{2loopxboxmore}.d). This also gives $6$ Riemann spheres.

Furthermore, combined with $K_1$, we get the topological picture
(\ref{2loopxboxmore}.e), which has $8$ Riemann spheres. (This case was
studied in \cite{Badger:2012dp}.) We computed
the geometric genera of all irreducible branches in this paragraph,
and found that they are all zero. Intersection points computation
from on-shell equations of irreducible branches, also agrees
with this analysis. This verifies the
result in \cite{Feng:2012bm}.
%

\section{Genera of algebraic curves from three-loop diagrams}

We also consider algebraic curves defined by equations of maximal
unitarity cut from 4-dimensional three-loop diagrams. There are
$4\times 3=12$ components in loop momenta, so diagrams with $11$
propagators would generate algebraic curves. We study the genera of
several three-loop curve examples. Although the algebraic system is more
complicated in these cases, the genus computation process is similar
to that of two-loop cases. We will project curves to plane
curves, and use the knowledge of arithmetic genus and singular points
to compute the geometric genera.

\subsection{Three-loop planar pentagon-box-box diagram, genus up to 1}

\begin{figure}
\center
  \includegraphics[width=6.2in]{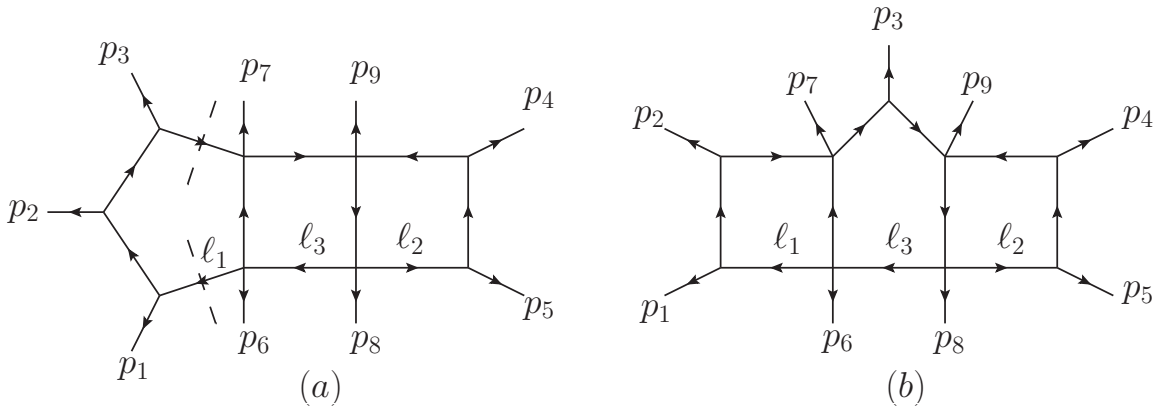}\\
  \caption{Planar three-loop diagrams with $11$ propagators: (a) pentagon-box-box diagram, (b) box-pentagon-box diagram.
  All external momenta are out-going and massive. The loop momenta are denoted
  by $\ell_1,\ell_2,\ell_3$.}\label{3loop}
\end{figure}

We consider the generic pentagon-box-box diagram with $9$ massive
external legs, as shown in Figure (\ref{3loop}.a).
The $11$ propagators are given by
\bea
&&D_0=\ell_1^2~,~D_1=(\ell_1-p_1)^2~,~D_2=(\ell_1-p_1-p_2)^2~,~D_3=(\ell_1-p_1-p_2-p_3)^2~,~~~\nonumber\\
&&\widetilde{D}_0=\ell_2^2~,~\widetilde{D}_1=(\ell_2-p_5)^2~,~\widetilde{D}_2=(\ell_2-p_4-p_5)^2~,~\bar{D}_0=\ell_3^2~,~~~\nonumber\\
&&\bar{D}_1=(\ell_3-p_1-p_2-p_3-p_6-p_7)^2~,~\widehat{D}_0=(\ell_3-\ell_1-p_6)^2~,~\widehat{D}_1=(\ell_3+\ell_2+p_8)^2~.~~~\nonumber\eea
The $11$ on-shell equations
$D_i=\widetilde{D}_i=\bar{D}_i=\widehat{D}_i=0$ of maximal unitarity
cut define an algebraic curve, and its topology is actually simple.
To see this, note that $4$ on-shell equations $D_i=0$ contain only
$\ell_1$, and we can solve them first. Since $\ell_1$ has $4$ components,
these $4$ equations will completely fix $\ell_1$. Explicitly, after
solving $3$ linear equations from $D_i=0$, the remaining one becomes
a quadratic equation of a single variable. This equation will give two
solutions for $\ell_1$, and on each solution, $\ell_1$ is completed
fixed. Hence $\ell_1$ can be treated as a $4$-vector similar to external
momenta. The remaining $11-4=7$ cut equations for $l_2$ and $l_3$
are the same as the two-loop double-box case.

More explicitly, for each solution of $\ell_1$, the on-shell
propagators highlighted by dashed lines in Figure (\ref{3loop}.a) can
be treated as massless external momenta. So the discussion reduces
to the two-loop double-box diagram with general kinematics if
$p_6,p_7$ are not absent. As we know it is a genus-$1$ torus. In some
specific kinematic configurations of $(p_4,p_5,p_6,p_7,p_8,p_9)$,
the algebraic curves are degenerate, and the topological pictures
are given by genus-$0$ Riemann spheres connected at points and
linked in a chain, just as the two-loop
double-box diagram. The two solutions of $\ell_1$ are separated.

\subsection{Three-loop planar box-pentagon-box diagram, genus up to 5}

On-shell equations of the three-loop box-pentagon-box diagram are much
more complicated. Naive elimination of linear equations will generate
$5$ quadratic equations in $6$ variables, and they define a
one-dimensional  curve. More explicitly, we consider
the generic box-pentagon-box diagram with $9$ massive external
momenta as shown in Figure (\ref{3loop}.b). Among the $11$
propagators, there are $3$ containing $\ell_1$ only
\begin{equation}
  \label{eq:8}
  D_0=\ell_1^2~,~D_1=(\ell_1-p_1)^2~,~D_2=(\ell_1-p_1-p_2)^2~,~~~
\end{equation}
$3$ containing $\ell_2$ only
\begin{equation}
 \widetilde{D}_0=\ell_2^2~,~\widetilde{D}_1=(\ell_2-p_5)^2~,~\widetilde{D}_2=(\ell_2 - p_4 -
 p_5)^2~,~~~
\end{equation}
and $3$ containing $\ell_3$ only
\begin{equation}
 \bar{D}_0=\ell_3^2~,~\bar{D}_1=(\ell_3 - p_1 - p_2 - p_6 -
 p_7)^2~,~\bar{D}_2=(\ell_3-p_1-p_2-p_3-p_6-p_7)^2~.~~~
\end{equation}
The remaining two propagators contain terms of mixed loop momenta,
and they are given by,
\bea
\widehat{D}_0=(\ell_3-\ell_1-p_6)^2~,~\widehat{D}_1=(\ell_2+\ell_3+p_8)^2~.~~~\eea
We have $4$ variables for each loop momentum, denoted as
$(x_1,x_2,x_3,x_4)$ for $\ell_1$, $(y_1,y_2,y_3,y_4)$ for $\ell_2$
and $(z_1,z_2,z_3,z_4)$ for $\ell_3$. As usual, for each loop
momentum we have $2$ linear equations, and by solving them, we can
eliminate $2$ variables and $2$ equations. So finally
we get $5$ quadratic polynomials,
\bea
&&Q_1(x_3,x_4)=D_0~,~Q_2(y_3,y_4)=\widetilde{D}_0~,~Q_3(z_3,z_4)=\bar{D}_0~,~~~\nonumber\\
&&Q_4(x_3,x_4,z_3,z_4)=\widehat{D}_0-D_0-\bar{D}_0~,~Q_5(y_3,y_4,z_3,z_4)=\widehat{D}_1-\widetilde{D}_0-\bar{D}_0~.~~~\eea
%

Notice that
$Q_1(x_3,x_4)=0$, $Q_2(y_3,y_4)=0$ and $Q_3(z_3,z_4)=0$ are conics. So
it is always possible to find rational parametrization for each loop
momentum, and express them by one free parameter as
$\ell_1=\ell_1(x)$, $\ell_2=\ell_2(y)$ and
$\ell_3=\ell_3(z)$\footnote{By coordinate transformation, we can
rewrite a non-degenerate conic $Q(x,y)=0$ in the form
$Q'(x',y')=x'y'-c=0$, and express one parameter as a rational function
of the other as $x'=c/y'$. So the loop momentum can be expanded as a
rational function of a single variable.}. The remaining two equations
$\widehat{D}_0$ and $\widehat{D}_1$ are meromorphic functions after
substituting $\ell_i$ back, and the numerators define an
algebraic plane curve as,
\begin{eqnarray}
 a_2(x) z^2+a_1(x) z+a_0(x) &=&0~,~~~ \nonumber \\
  b_2(y) z^2+b_1(y) z+b_0(y)&=&0~.~~~
 \label{box-penta-box}
\end{eqnarray}
For generic massive external momenta, $a_i$ and $b_i$ are quadratic
polynomials. We can eliminate $z$ by computing the {\it resultant}
of these two equations to get a plane curve $F(x,y)$=0. This plane
curve is birationally equivalent to (\ref{box-penta-box}) via the
inverse map,
\begin{equation}
  \label{InVTransform}
  z=\frac{-a_2(x) b_0(y)+a_0(x) b_2(y)}{a_2(x) b_1(y)-a_1(x)
  b_2(y)}~,~~~
\end{equation}
so they have the same geometric genus. For
generic kinematic configurations, the plane curve $F(x,y)$ has the degree
$8$, so the arithmetic genus is
\bea g_A={(d-1)(d-2)\over 2}=21~.~~~\eea
There are $4$ singular points of the multiplicity $\mu_P=2$, and $2$
singular points of the multiplicity $\mu_P=4$, so the geometric genus is
given by
\begin{equation}
  \label{eq:11}
  g_G=g_A-\sum_{P\in Sing(P)}{1\over 2}\mu_P(\mu_P-1)=21-4\times
  1-2\times 6=5~.~~~
\end{equation}
So this is a genus 5 curve.

Alternatively, we can also find the genus of (\ref{box-penta-box})
by Riemann-Hurwitz formula (\ref{Riemann-Hurwitz})\footnote{We thank
  Simon Caron-Huot for the discussion of using this formula.}. Let $\mathcal C$ be
the complex curve defined by (\ref{box-penta-box}). The projection $f:
(x,y,z)\mapsto z$ is a covering map from $\mathcal C$ to the complex
plane of $z$. For all but finite points, this is a four-fold covering
so the degree of the map is $4$. We determine the ramified points as
follows:
\begin{enumerate}
\item Consider the first equation in (\ref{box-penta-box}) as a
  quadratic equation for $x$ and the discriminant is
  $\Delta_1(z)$. For a fixed value of $z$, there are two corresponding
  $x$ unless $\Delta_1(z)$ vanishes. We can checked that $\Delta_1(z)=0$ have
  four distinct roots on the $z$ plane, namely $z_1$, $z_2$, $z_3$ and
  $z_4$.
\item Consider the second equation in (\ref{box-penta-box}) as a
  quadratic equation for $y$ and the discriminant is
  $\Delta_2(z)$. We can checked that $\Delta_2(z)=0$ have
  four distinct roots on the $z$ plane, namely $z_5$, $z_6$, $z_7$ and
  $z_8$.
\item We checked $z_i$'s are all disctint, $i=1,\ldots, 8$. Hence for
  each $z_i$, $i=1,\ldots 4$, there are two corresponding points on
  $\mathcal C$, namely $(x_i,y_i^{(1)},z_i)$ and
  $(x_i,y_i^{(2)},z_i)$. For
  each $z_i$, $i=5,\ldots 8$, there are two corresponding points on
  $\mathcal C$, namely $(x_i^{(1)},y_i,z_i)$ and
  $(x_i^{(2)},y_i,z_i)$. So there are $2\times 4+2\times 4=16$
  ramified points on  $\mathcal C$ and each point has the ramification
  index $2$.
\end{enumerate}
Then we can use Riemann-Hurwitz formula (\ref{Riemann-Hurwitz}) and
the fact that the complex plane has the genus zero,
\begin{equation}
  \label{eq:12}
  2 g(\mathcal C)-2=4\times (2\times 0-2)+16\cdot (2-1)~,~~~
\end{equation}
Again we get the same conclusion that the curve has the genus $5$.

Finally, we can also find the genus with the help of two-loop-diagram genus
information. If we forget the cut equations involves $l_2$ for a
while, the rest equations form the hepta-cut equations of the two-loop massive
double box diagram. Intuitively, it is clear that if all propagators
involving $l_2$ are pinched, the resulting diagram is the massive
double box. Explicitly, neglect the second equation
in (\ref{box-penta-box}) and we find that the remaining equation,
\begin{eqnarray}
 a_2(x) z^2+a_1(x) z+a_0(x) &=&0~,~~~
\end{eqnarray}
which corresponds to the massive double box topology, defines an
elliptic curve with the genus $1$. Then we consider the projection of
the variety (\ref{box-penta-box}) to this elliptic curve, via
\begin{equation}
  \label{eq:23}
  (x,y,z) \mapsto (x,z)
\end{equation}
This is a ramified double covering. The ramified point are determined
by the discriminant $\Delta(z)$ for the equation,
\begin{eqnarray}
 b_2(y) z^2+b_1(y) z+b_0(y) &=&0~,~~~
\end{eqnarray}
in $y$. The simultaneous equations $\Delta(z)=0$ and $a_2(x) z^2+a_1(x)
z+a_0(x)=0$ has $8$ distinct solutions. Then we can use Riemann-Hurwitz formula (\ref{Riemann-Hurwitz}) and
the fact that the complex plane has the genus zero,
\begin{equation}
  \label{eq:12}
  2 g(\mathcal C)-2=2\times (2\times 1-2)+8\cdot (2-1)~,~~~
\end{equation}
As expected, we get the same conclusion that the curve has the genus
$5$. Note that here we explicitly used the genus information of the
massive double-box topology. In this sense,  Riemann-Hurwitz formula
provides an induction relation between the three-loop box-penta-box
diagram and two-loop double box diagram.

\subsubsection{Degeneracy under specific kinematics}

The degeneracy pattern of the genus-$5$ Riemann surface is more complicated
than that of genus-$3$ Riemann surface of two-loop example. 
Here we will discuss some degenerate cases.

If at least one  of $p_1,p_2$ is massless, $Q_1(x_3,x_4)$ is
factorized and there are two irreducible branches.  Similarly, if at least one
of $p_4,p_5$ is massless, $Q_2(y_3,y_4)$ has two
factors. Also, if $p_3$ is massless, $Q_3(z_3,z_4)$ has
two factors. When one of $p_6,p_7,p_8,p_9$ is
absent, the algebraic set defined by $Q_i,i=1,..,5$ is also reducible, but
the factorization is not obvious. Again, we
can get two irreducible branches from primary decomposition method by
Macaulay2 \cite{M2}.

\begin{figure}
\center
  \includegraphics[width=6in]{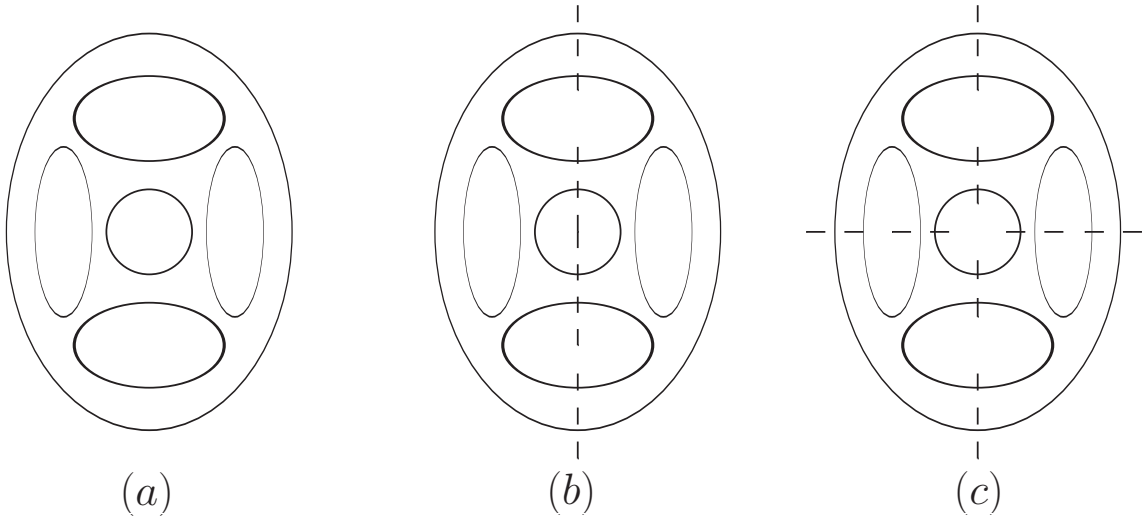}\\
  \caption{Topological pictures of degenerate on-shell equations from planar three-loop
  box-pentagon-box diagram under
  specific kinematic configurations. (a) the curve is irreducible, and
  the genus is 5, (b) the curve has 2 irreducible branches, each branch
  has genus 1. They are connected by 4 points, (c) the curve
  has 4 irreducible branches, each branch has genus 0. They are
  linked in chain with two adjacent branches connected by 2 points. }\label{boxpentaboxgenus}
\end{figure}

Let us assume $p_1^2=0$, and the quadratic polynomial of $\ell_1$
has two factors $Q_1(x_3,x_4)=f_1(x_3,x_4)f_2(x_3,x_4)$, where
$f_1,f_2$ are both linear in $x_3,x_4$. The two irreducible branches
are defined by $I_1=\langle f_1,Q_2,Q_3,Q_4,Q_5\rangle$ and
$I_2=\langle f_2,Q_2,Q_3,Q_4,Q_5\rangle$. By the elimination  method
via Gr\"obner basis, we can get a plane curve of degree $8$ for each
branch, so the arithmetic genus is $g_A=21$. There are $8$ singular
points of the multiplicity $\mu_P=2$, and $2$ singular points of the
multiplicity $\mu_P=4$, so the geometric genus is $g_G=21-8\times
1-2\times 6=1$. We can also count the intersection points between
two curves, and there are $4$ single points. So the topological
picture of this degeneracy can be given by contracting tubes to
points along dashed lines, as shown in Figure
(\ref{boxpentaboxgenus}.b).
%

We consider another kinematic configuration where $p_1$ and
$p_5$ are massless. In this case both $Q_1(x_3,x_4)$ and
$Q_2(y_3,y_4)$ are degenerate, and we can get $4$ irreducible
branches. From Macaulay2, we can also get $4$ irreducible ideals. For
each branch, we can get a plane curve of degree $4$ via Gr\"obner
basis method, so the arithmetic genus is $g_A=3$. There are $3$
singular points of the multiplicity $\mu_P=2$, so the geometric genus is
$g_G=3-3\times 1=0$. Counting intersecting points, we found that
there are $8$ points in total: the $4$ irreducible branches are
linked in chain, and the adjacent two branches intersect at $2$
points. The topological picture can be given by contracting
tubes to points around dashed lines, as shown in Figure
(\ref{boxpentaboxgenus}.c). Each branch is a genus-$0$
Riemann sphere.

\subsection{Three-loop non-planar box-crossed-pentagon diagram, genus
  up to 9}

\begin{figure}
\center
  \includegraphics[width=6in]{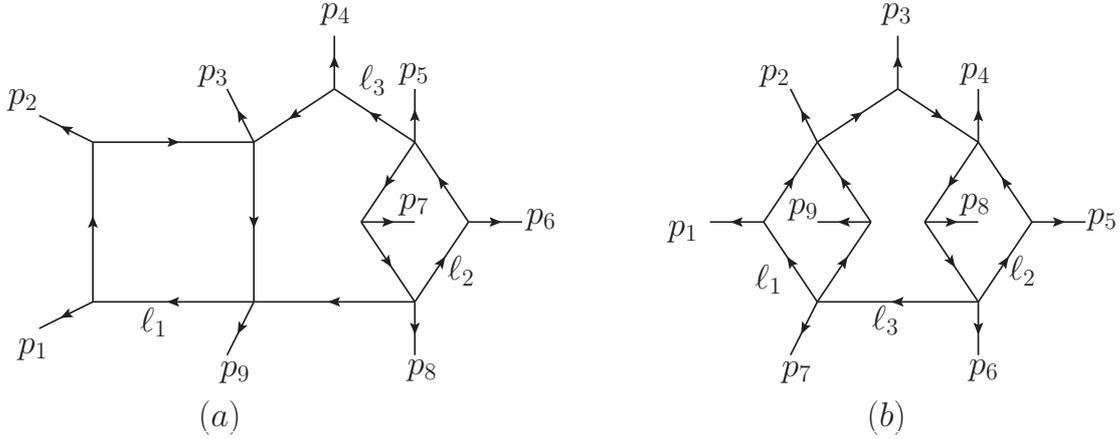}\\
  \caption{(a) Non-planar three-loop box-crossed-pentagon
  diagram,(b) Non-planar three-loop crossed-crossed-pentagon
  diagram.
  All external momenta are out-going and massive. The loop momenta are denoted
  by $\ell_1,\ell_2,\ell_3$.}\label{3loopboxcrossedpentagon}
\end{figure}

In this subsection, we discuss the three-loop box-crossed-pentagon
diagram as shown in Figure (\ref{3loopboxcrossedpentagon}.a).
This is a non-planar three-loop diagram with $9$ general massive
external momenta. As usual, the maximal unitarity cut of this
diagram gives $11$ on-shell equations of $12$ variables, and they
define an algebraic curve. Again, there are $3$ propagators
containing only $\ell_1$,
\begin{equation}
  \label{eq:8}
  D_0=\ell_1^2~,~D_1=(\ell_1-p_1)^2~,~D_2=(\ell_1-p_1-p_2)^2~.~~~
\end{equation}
From on-shell equations $D_i=0$ we can parameterize $\ell_1$
rationally by one free parameter $x$. Similarly, there are $3$
propagators containing only $\ell_3$,
\begin{equation}
 \bar{D}_0=\ell_3^2~,~\bar{D}_1=(\ell_3 - p_4)^2~,~\bar{D}=(\ell_3 - p_1 - p_2 - p_3 -
 p_4-p_9)^2~.~~~
\end{equation}
Using on-shell equations $\bar{D}_i=0$, $\ell_3$ is rationally
parameterized by one free parameter $w$. However, only $2$
propagators
\begin{equation}
  \label{eq:13}
  \widetilde{D}_0=\ell_2^2~,~\widetilde{D}_1=(\ell_2-p_6)^2~~~
\end{equation}
containing single loop momentum $\ell_2$, so $\ell_2$ is rationally
parameterized by $2$ free parameters, namely, $y$ and $z$ from
on-shell equations $\widetilde{D}_i=0$.

The remaining $3$ propagators contain terms of mixed loop momenta,
\bea
  \label{eq:14}
 &&\widehat{D}_0=(\ell_2-\ell_3-p_5-p_6)^2~,~\widehat{D}_1=(\ell_2-\ell_3-p_5-p_6-p_7)^2~,~\nonumber\\
 &&\widehat{D}_2=(\ell_1+\ell_3-p_1-p_2-p_3-p_4)^2~.~~\eea
After substituting $\ell_1(x),\ell_2(y,z),\ell_3(w)$ back, they
become meromorphic functions. The numerators $f_1,f_2,f_3$ of these
$3$ meromorphic functions are polynomials in $(x,y,z,w)$, and
equations
\begin{equation}
  \label{eq:15}
  f_1(y,z,w)=f_2(y,z,w)=f_3(x,w)=0
\end{equation}
define the algebraic curve. They are not necessary quadratic.

Our strategy is to eliminate $y$ and $w$ from the equations and then
get a plane curve in $x$ and $z$. This can be done automatically by
Gr\"obner basis method. However, it is helpful to eliminate $y$ and
$w$ step by step, so that we can explicitly show this projection process is
birational. Furthermore, we can see the induction relation from the
two-loop non-planer diagram to this diagram.

First, note that if we combine the external legs $p_1$, $p_2$, $p_3$
and $p_9$ to one external leg and neglect the four cut equations
involving $l_1$, then we get the prime case of two-loop non-planar
crossed-box diagram. In other words, $f_1(y,z,w)=f_2(y,z,w)=0$
defines a genus-3 complex curve. Hence we can eliminate $y$ like the
case (\ref{box-penta-box}) and it is birational via an inverse
transformation like (\ref{InVTransform}). The resulting reduced
algebraic system is
\begin{gather}
a_8w^8+a_7(z) w^7+a_6(z) w^6+a_5(z) w^5+a_4(z) w^4 +a_3(z) w^3+ a_2(z) w^2+a_1(z) w+a_0(z) =0~,~~~ \nonumber \\
b_2(x) w^2+b_1(x) w+b_0(x)=0~,~~~ \label{box-xpenta}\end{gather}
where for generic massive external momenta, $a_i$, $i=0,\ldots 6$ and $b_i$, $i=0,\ldots 2$ are
quadratic polynomials while $a_7(z)$ is linear and $a_8$ is a
constant.  Again, we use the resultant to eliminate $w$
and get a plane curve $F(x,z)=0$. This step is also birational,
and the inverse map is
\begin{equation}
  \label{eq:16}
  w= \frac{p(x,z)}{q(x,z)}~,~~~
\end{equation}
where the explicit expressions for $p(x,z)$ and $q(x,z)$ can be found by Gr\"obner basis method.

Finally, the plane curve $F(x,z)=0$ has the degree $20$, so the
arithmetic genus is $g_A=171$. There are $32$ normal singular points
of the multiplicity $\mu_P=2$, one normal singular point of the
multiplicity $\mu_P=4$, and one $16$-fold point. The $16$-fold point
has $8$ ordinary tangent lines with the multiplicity $1$ and $4$
tangent lines with the multiplicity $2$. Since this $16$-fold point
is not a normal singular point, we need to perform a blow up. The
new blow-up curve resolves the $16$-fold point into $8$ smooth
points and $4$ normal double points. In summary,  the genus is
\begin{equation}
  \label{eq:17}
  g_G=\frac{1}{2}(20-1)(20-2)-32\cdot 1-\frac{1}{2}\cdot 4\cdot (4-1)-\frac{1}{2}\cdot 16\cdot
  (15-1)-4=9~.~~~
\end{equation}
So the maximal unitarity cut of the non-planar three-loop
box-crossed-pentagon diagram, with all massive legs, defines a one-dimensional algebraic
curve of the genus 9.

Alternatively, we can use Riemann-Hurwitz formula to get the genus
of the curve (\ref{box-xpenta}). The first equation of
(\ref{box-xpenta}) generates $8$ ramified points on the $w$ plane
while the second equation generates $4$ ramified points on the $w$
plane. All these points are distinct, so there are $8\cdot 2+4\cdot
2=24$ ramified points on the curve (\ref{box-xpenta}). So
Riemann-Hurwitz formula reads
\begin{equation}
  \label{eq:10}
  2 g-2=4 (0-2)+24~,~~~
\end{equation}
which again implies that (\ref{box-xpenta}) has the genus $9$.

\subsubsection{Degeneracy under specific kinematics}

Before going to the discussion of degenerate cases, let us first
verify an observation of on-shell equations\footnote{We thank Simon
Caron-Huot for this observation.}. If we forget the $4$ equations
that containing $\ell_1$, i.e., $D_0=D_1=D_2=\widehat{D}_2=0$, then
the remaining $7$ equations describe the prime case of two-loop
non-planar crossed-box diagram, so it should have the genus 3. To see
this, we could first eliminate $4$ variables from $4$ linear
equations $\bar{D}_0-\bar{D}_1=\bar{D}_0-\bar{D}_2=0$,
$\widetilde{D}_0-\widetilde{D}_1=0$ and
$\widehat{D}_0-\bar{D}_0-\widetilde{D}_0=0$. Then the remaining $3$
equations of $4$ variables define an equivalent curve, and we can
birationally project it to a plane curve. The resulting plane curve
has the degree $8$, while there are $12$ normal singular points of the
multiplicity $m_P=2$, $1$ normal singular point of the multiplicity
$m_P=4$, so the genus is
\bea g_G={1\over 2}(8-1)(8-2)-12\cdot 1-{1\over 2}4(4-1)=3~,~~~\eea
which agrees with the observation.

The genus-9 curve of generic kinematics is complicated. However, as
previous examples, we can still consider some kinematic limits where
on-shell equations are degenerated. In these cases the solution
space has several branches, and each branch defines a curve with the
lower genus. Here we will show some simple degenerate cases.
\begin{figure}
\center
  \includegraphics[width=6in]{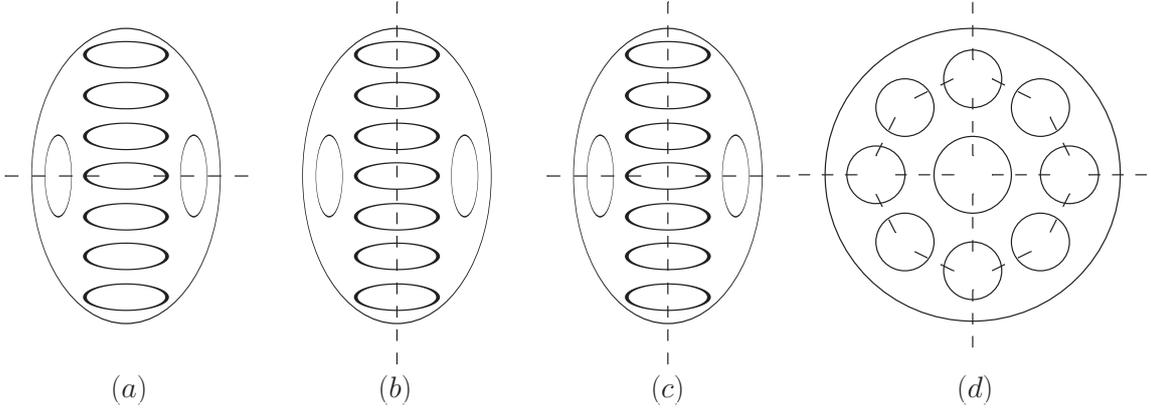}\\
  \caption{Topological pictures of on-shell equations from the non-planar three-loop
  box-crossed-pentagon diagram under
  specific kinematic configurations. (a) the curve has 2 irreducible
  branches, each branch has the genus 3 and they are connected
  by 4 points, (b) the curve has 2 irreducible branches, each
branch has the genus 1, and they are connected by 8 single points, (c)
the curve
  has 4 irreducible branches, each branch has the genus 0, and they are
  linked in chain, (d) the curve has 8 irreducible branches, and each
  branch has the genus 0. }\label{boxXpentagongenus}
\end{figure}

After solving linear equations, we get $5$ quadratic equations,
\bea
&&Q_1(x_1,x_2)=0~,~Q_2(y_1,y_2,z_1,z_2)=0~,~Q_3(z_1,z_2)=0~,~~~\nonumber\\
&&Q_4(x_1,x_2,z_1,z_2)=0~,~Q_5(y_1,y_2,z_1,z_2)=0~,~~~\eea
where $Q_1,Q_2,Q_3$ come from $\ell_1^2,\ell_2^2,\ell_3^2$ and
$Q_4,Q_5$ come from the mixed terms $\ell_1\ell_3, \ell_2\ell_3$. If
$p_1$ or $p_2$ is massless, then $Q_1$ has two
factors, while if $p_4$ is massless, then $Q_3$ has
two factors. In these kinematic configurations, there are two
irreducible branches. As usual, we can project each branch onto a plane curve. The resulting plane curve has
the degree $12$, so the arithmetic genus is $g_A=55$. There are $16$
normal singular points of the multiplicity $m_P=2$, $1$ normal singular
point of the multiplicity $m_P=4$, one $8$-fold singular point. This
$8$-fold point is not normal, and after blowing up, we get $2$
normal singular points of the multiplicity $m_P=2$. So finally we get,
\bea g_G=55-16\cdot 1-{1\over 2}4(4-1)-{1\over 2}8(8-1)-2\cdot
1=3~.~~~\eea
Therefore there are two genus-3 curves. We can also calculate the
intersecting points between these two curves, and find that there are $4$ points. The topological picture can be
reproduced from the genus-9 picture by contracting tubes to points
along dashed lines as shown in Figure
(\ref{boxXpentagongenus}.a).

If $p_6$ or $p_7$ is massless, $Q_2$ has two
factors. In this case, again, there are two irreducible branches.
Similarly, after birationally projecting each
branch onto a plane curve, we get an equation of the degree $12$. There
are $20$ normal singular points of the multiplicity $m_P=2$, one normal
singular point of the multiplicity $m_P=4$ and one normal singular point
of the multiplicity $m_P=8$, so finally we get genus
\bea g_G={1\over 2}(12-1)(12-2)-20\cdot 1-{1\over 2}4(4-1)-{1\over
2}8(8-1)=1~.~~~\eea
There are $8$ intersection points between the two irreducible branches
. So the topological
picture can be reproduced from the genus-9 picture by contracting tubes
to points along dashed lines, as shown in Figure
(\ref{boxXpentagongenus}.b).

In fact, 
if we know the genus for generic kinematics, together with the
number of irreducible branches for specific kinematic configuration
and the intersecting points between different branches, it is
possible to predict the genus of each branch by directly studying the
topological picture. Take
the kinematic configuration in previous paragraph as an example:
intuitively, these two branches are symmetric, and they should have the
same genus. A simple calculation determined that there are $8$
intersecting points. Then it is clear that    (\ref{boxXpentagongenus}.b) is the only possible
topological picture which satisfies these requirements.  So
we can conclude that each branch is genus-$1$ without doing any further
calculation.

If both $p_1$ and $p_6$ are massless, $Q_1$ and $Q_2$ are factorized, and each of them has two factors. Thus, the variety has
$4$ irreducible branches. The on-shell equation system of each branch is
quite simple, and the corresponding plane curve for each branch has
the
degree $4$. There are $3$ normal singular points of the multiplicity
$m_P=2$, so the genus is $g_G=(4-1)(4-2)/2-3=0$. The topological
picture is shown in Figure (\ref{boxXpentagongenus}.c).

 If we
further consider the case with massless $p_1,p_4,p_6$, there are
$8$ irreducible branches. The plane curve of each branch has the degree 3,
and there is only one normal singular point of the multiplicity $m_P=2$,
so the genus $g_G=(3-1)(3-2)/2-1=0$. After obtaining the
intersecting points between branches, we draw the topological
picture in Figure (\ref{boxXpentagongenus}.d).

\subsection{Three-loop non-planar crossed-crossed-pentagon diagram,
  genus up to 13}

The same methods can be used to study the non-planar three-loop
crossed-crossed-pentagon diagram, as shown in Figure
(\ref{3loopboxcrossedpentagon}.b), although the computation is more
complicated. Nevertheless we should start from the on-shell
equations. There are $2$ propagators containing only $\ell_1$
\begin{equation}
  D_0=\ell_1^2~,~D_1=(\ell_1-p_1)^2~,~~~
\end{equation}
$2$ propagators containing only $\ell_2$
\begin{equation}
  \bar{D}_0=\ell_2^2~,~\bar{D}_1=(\ell_2-p_5)^2~,~~~
\end{equation}
and $3$ propagators containing only $\ell_3$
\begin{equation}
  \widetilde{D}_{0}=\ell_3^2~,~\widetilde{D}_1=(\ell_3-p_1-p_2-p_7-p_9)^2~,~\widetilde{D}_{2}=(\ell_3+p_4+p_5+p_6+p_8)^2~.~~~
\end{equation}
From maximal unitarity cut, we can rationally parameterize $\ell_1$
by two parameters $x,y$, $\ell_2$ by $z,w$, and $\ell_3$ by $\tau$.
The remaining $4$ on-shell equations
$\widehat{D}_0=\widehat{D}_1=\widehat{D}_2=\widehat{D}_3=0$, where
\bea
  \label{eq:18}
&&  \widehat{D}_0=(\ell_3-\ell_1-p_7)^2~,~\widehat{D}_1=(\ell_3-\ell_1-p_7-p_9)^2~,~~~\nonumber \\
&&\widehat{D}_2=(\ell_3+\ell_2+p_6)^2~,~\widehat{D}_3=(\ell_3+\ell_2+p_6+p_8)^2~,~~~\eea
become
\bea
   f_1(x,y,\tau)=f_2(x,y,\tau)=f_3(z,w,\tau)=f_4(z,w,\tau)=0~,~~~
\eea
which are polynomial equations. Again, it is clear that this diagram
contains two copies of non-planar two-loop diagrams. In other words,
$f_1=f_2=0$ defines a genus-3 curve and $f_3=f_4=0$ defines another
genus-3 curve. Following the previous analysis, we can birationally
eliminate $y$ and $w$ and get,
\bea
  g_1(x,\tau)=0~,~g_2(z,\tau)=0~,~~~ \label{xxpenta} \eea
where $g_1$ is quadratic in $x$ and $g_2$ is quadratic in $z$.

We can further eliminate $\tau$ to get a plane curve $F(x,z)=0$ in
$z$. The plane curve has the degree $24$, so the
arithmetic genus is $g_A=253$. There are $184$ normal singular points
of the multiplicity $\mu_P=2$, two normal singular point of the multiplicity
$\mu_P=8$. In summary, the genus is
\begin{equation}
  \label{eq:17}
  g_G=\frac{1}{2}(24-1)(24-2)-184\cdot 1-\frac{1}{2}\cdot 8\cdot
  (8-1)\cdot 2=13~.~~~
\end{equation}

Alternatively, each of $g_1$ and $g_2$ generates $8$
ramified points on the $\tau$ plane. All these points are distinct,
so there $8\cdot 2+8\cdot 2=32$ ramified points on the curve
(\ref{xxpenta}). So Riemann-Hurwitz formula reads
\begin{equation}
  \label{eq:10}
  2 g-2=4 (0-2)+32~,~~~
\end{equation}
which again implies that the curve from crossed-crossed-pentagon
diagram, with all massive legs, has the genus $13$.

\subsubsection{Degeneracy under specific kinematics}

Again, if we forget the $4$ equations containing $\ell_1$(or
$\ell_2$), as in the box-crossed-pentagon example, the remaining $7$
equations also describe the generic two-loop non-planar crossed-box
diagram, thus it should be genus 3. This can be verified by the same
method in the previous subsection.
%
%
%
\begin{figure}
  \includegraphics[width=6in]{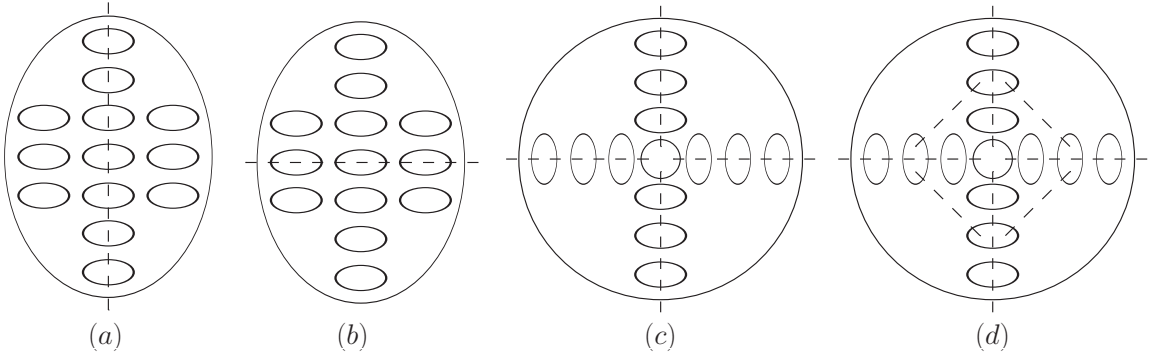}\\
  \caption{Topological pictures of on-shell equations from
    the non-planar three-loop
crossed-crossed-pentagon diagram under
  specific kinematic configurations. (a) The curve has 2 irreducible
  branches, and each branch has the genus 3. They are connected
  by 8 points, (b) The curve has 2 irreducible branches, and each
branch has the genus 5. They are connected by 4 points, (c)
The curve
  has 4 irreducible branches, and each branch has the genus 0. They are
  linked in chain with adjacent branches intersecting at 4 points,
  (d) The curve has 8 irreducible branches, and each branch has the genus 0.}\label{XXpentagongenus}
\end{figure}

We discuss some kinematic configurations, where the on-shell
equations are degenerated. Again, after solving linear equations
$D_0-D_1=0$ and $\widehat{D}_0-\widetilde{D}_0-D_0=0$, we can
formally write the other two quadratic equations as
\bea Q_1(x_1,x_2,z_1,z_2)=0~,~Q_4(x_1,x_2,z_1,z_2)=0~.~~~\eea
If $p_1$ is massless, $Q_1$ has two factors $Q_1=f_1f_2$, where
$f_1,f_2$ are linear functions. Thus, there are two irreducible
branches, from ideals $I_1=\langle f_1,Q_2,Q_3,Q_4,Q_5\rangle$ and
$I_2=\langle f_2,Q_2,Q_3,Q_4,Q_5\rangle$. These two branches
intersect at $8$ points, so we can guess that the topological
picture should be Figure (\ref{XXpentagongenus}.a) and each branch
gives a curve with the genus $3$. To verify this, as usual, we can
project the curve onto a plane curve, and the resulting equation has
the degree $12$. There are $40$ normal singular points of the
multiplicity $m_P=2$, $2$ normal singular points of the multiplicity
$m_P=4$, so the genus is
\bea g_G={1\over 2}(12-1)(12-2)-40\cdot 1-2\cdot {1\over
2}4(4-1)=3~,~~~\eea
which is exactly the result shown in Figure
(\ref{XXpentagongenus}.a).

If $p_3$ is massless, $Q_3$ has two factors.
There are two irreducible branches. The corresponding plane curve
of each branch has the degree $12$. And there
are $38$ normal singular points of the multiplicity $m_P=2$, $2$ normal
singular points of the multiplicity $m_P=4$, so the genus is given by
\bea g_G={1\over 2}(12-1)(12-2)-38\cdot 1-2\cdot {1\over
2}4(4-1)=5~.~~~\eea
The number of intersecting points between two branches is 4, so the
topological picture is given by contracting tubes to points from
the genus-$13$ picture, as shown in Figure (\ref{XXpentagongenus}.b).

It is not hard to predict that if both $p_1,p_3$ are massless, the
topological picture should be given by overlapping of Figure
(\ref{XXpentagongenus}.a) and (\ref{XXpentagongenus}.b). In other
words, there are $4$
genus-$1$ tori, which are linked in a chain. In fact, in this kinematic
configuration, both $Q_1$ and $Q_3$ are factorized, and there are
indeed $4$
branches. The plane curve of each branch has the degree $6$, so the
arithmetic genus is $g_A=(6-1)(6-2)/2=10$. There are $9$
normal singular points of the multiplicity $m_P=2$, so the genus from
explicit calculation is
\bea g_G=10-9\cdot 1=1~,~~~\eea
as expected.

If both $p_1,p_5$ are massless, $Q_1,Q_2$ are factorized. So there
are $4$ irreducible branches. The plane curve of each branch has the
degree $6$, while there are $10$ normal singular points of the
multiplicity $m_P=2$, so the genus is $0$. If $p_1,p_3,p_5$ are
massless, we further get $8$ branches. The plane curve of each
branch has the degree $3$, and there is only one singular point of
the multiplicity $1$, so the genus is again $0$. After obtaining the
intersecting points between branches, we can sketch the topological
pictures of both kinematic configurations from the genus-$13$
picture in Figure (\ref{XXpentagongenus}.c) and
(\ref{XXpentagongenus}.d).

\subsection{Three-loop Mercedes-logo diagram, genus up to $9$}

So far, we only consider the three-loop diagrams with ``ladder''
type. The same mathematical method also works for three-loop
diagrams with ``Mercedes-logo'' topology.
\begin{figure}
\center
  \includegraphics[width=3in]{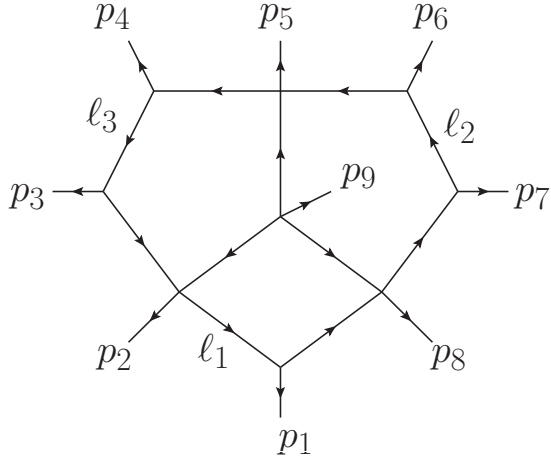}\\
  \caption{Non-planar three-loop "Mercedes" logo diagram.
  All external momenta are out-going and massive. The loop momenta are denoted
  by $\ell_1,\ell_2,\ell_3$.}\label{3loopMercedes}
\end{figure}

In this subsection, we explicitly calculate a three-loop
``Mercedes-logo'' diagram with $11$ propagators and $9$ massive
external legs as shown in Figure (\ref{3loopMercedes}). Different
from the "ladder" diagrams, all possible terms of loop momenta
$\ell_1\ell_2,\ell_1\ell_3,\ell_2\ell_3$ will appear. This will
somehow complicate the cut equations. There are two propagators
containing only $\ell_1$
\bea D_0=\ell_1^2~~,~~D_1=(\ell_1-p_1)^2~,~~~\eea
and three propagators containing only $\ell_2$
\bea
\bar{D}_0=\ell_2^2~~,~~\bar{D}_1=(\ell_2-p_6)^2~~,~~\bar{D}_2=(\ell_2+p_7)^2~,~~~\eea
and three propagators containing only $\ell_3$
\bea
\widetilde{D}_0=\ell_3^2~~,~~\widetilde{D}_1=(\ell_3-p_3)^2~~,~~\widetilde{D}_2=(\ell_3+p_4)^2~.~~~\eea
The remaining three propagators contain mixed terms as
\bea &&\widehat{D}_0=(\ell_2 - \ell_1+ p_7+ p_1 +
p_8)^2~~,~~\widehat{D}_1=(\ell_3 - \ell_2+ p_4  +
 p_6 + p_5)^2~,~~~\\
 &&\widehat{D}_2=(\ell_1 - \ell_3 + p_3 + p_2)^2~.~~~\eea
It is not possible to construct linear equations directly from
equations of mixed propagators, so we can only get 5 linear
equations from $D_i=\bar{D}_i=\widetilde{D}_i=0$. The
one-dimensional curve is described by 6 quadratic equations of 7
variables.

As before, we can parameterize $\ell_1$ by two variables $x$ and
$y$, $\ell_2$ by one variable $z$, and $\ell_3$ by one variables
$w$. The rest equations have the following form
\begin{eqnarray}
  f(x,y,z)=0~~,~~g(w,z)=0~~,~~h(x,y,w)=0~.~~~
\end{eqnarray}

We birationally project the variety on a plane curve $F(y,w)=0$.
This plane curve has the degree $20$, $36$ normal double point, one
normal $16$-fold point and one normal $4$-fold point. So the
geometric genus is,
\begin{equation}
  g_G=\frac{(20-1)(20-2)}{2}-36-\frac{16(16-1)}{2}-\frac{4(4-1)}{2}=9~.~~~
\end{equation}

Interestingly, genus of Mercedes logo diagrams equals to the genus
of box-crossed-pentagon diagrams, they both have genus 9. We examine
some simple degenerate equations under specific kinematic
configurations, and find that the topological pictures are the same
as box-crossed-pentagon. If $p_4$ is massless, there are two
irreducible branches. After birationally projecting each branch onto
plane curve, we get an equation of degree 12. There are 18 normal
singular points of $m_P=2$, 1 normal singular point of $m_P=4$ and 1
normal singular point of $m_P=8$, so the genus is 3. These two
branches intersect at 4 points, and the topological picture is the
same as shown in Figure (\ref{boxXpentagongenus}.a). If $p_1$ is
massless, cut equations also degenerate to two irreducible branches.
The plane curve of each branch has degree 12. There are 20 normal
singular points of $m_P=2$, 1 singular point of $m_P=4$ and 1
singular point of $m_P=8$, so the genus is 1. There are 8
intersecting points between two branches, and the picture is the
same as Figure (\ref{boxXpentagongenus}.b). If both $p_1,p_4$ are
massless, there are 4 irreducible branches. The plane curve after
birationally projecting each branch has degree 6 and 10 normal
singular points, so each of them is genus-0 curve. The topological
picture is the same as Figure (\ref{boxXpentagongenus}.c). Further
considering $p_1,p_4,p_6$ massless, we get 8 irreducible branches.
Each branch can be projected onto a plane curve of degree 3 and 1
singular point, so they are again genus-0 curve. After getting the
intersecting points among branches, the topological picture can be
found to be the same as the one shown in Figure
(\ref{boxXpentagongenus}.d).

\section{Conclusion}

In this paper, we study the global structure of on-shell equations
from generalized unitarity cut of loop diagrams. We focus on the
$L$-loop diagram with $4L-1$ unitarity cuts in four-dimensional theories.
There is one degree of freedom for the solution space, which is a
complex algebraic curve or union of complex algebraic curves.
Since the topology of a complex algebraic curve is
completely determined by the genus, in order to study the global
structure, we compute the genera of algebraic curves. This is systematically done by computational
algebraic geometry methods.


In this paper, we used two algebraic geometry methods to determine the
genera of curves from unitarity cuts: (1) the relation between {\it
arithmetic genus} and {\it geometric genus} (2) Riemann-Hurwitz
formula. These methods clearly work for all
algebraic curves, and are quite efficient for those from two and three-loop unitarity cuts.

 The only algebraic curve for the one-loop case is from the one-loop triangle
diagram with $3$ propagators. The defining equation is simply a conic
section, and it is equivalent to a genus-$0$ Riemann sphere, so we can always
find the rational parametrization for the solution space.

Two-loop diagrams with $7$ propagators also define algebraic curves.
We find that the genus could be as high as $3$, in the two-loop
non-planar crossed-box diagram with the maximal number of massive external
momenta. In the degenerate limit, i.e., some external momenta being
massless or absent, the solution space degenerates to the union of
several complex curves. These curves have lower genera, $1$ or $0$.
As the global structure analysis in \cite{CaronHuot:2012ab}, from
intersection structure of these curves, we can show that the union
of these curves exactly comes from a degenerate genus-$3$ curve.

The topology of algebraic curves from three-loop diagrams with $11$
propagators is more complicated. We consider four examples.
The pentagon-box-box diagram is, in some sense,   similar to the two-loop
double box diagram. The curve from the planar box-pentagon-box diagram with
generic kinematics has the genus $5$. For the non-planar box-crossed-pentagon
diagram with generic kinematics, the curve can have the genus $9$. And
for the  non-planar crossed-crossed-pentagon diagrams, the curve can
have the genus $13$. Again, for diagrams with massless external legs, we
show that the global structure of solutions comes from a degenerate high-genus curve.

The information of genus and the topological picture under
degenerate limit is important for calculating multi-loop
amplitude via unitarity cut method:
\begin{itemize}
\item The genus is the criterion for
rational parametrization: if the curve has the genus zero, then it can
be rationally parameterized. Otherwise, the rational parametrization
does not exist. This property is important, since for unitarity
computation, we may
parameterize solution space before tree amplitude computations. So it
is useful to know the difficulty of parametrization before unitarity
computation.
\item The
topological picture under degenerate limit is important
for calculating
expansion coefficients of integrand basis. Usually, all branches
should be studied to get these coefficients. Knowing the genus for
generic kinematics, it is possible to predict the topology for
branches under degenerate limits from the knowledge of
intersection pattern. Especially, we can use the genus information to
check if all the cut solutions are identified or not.
\end{itemize}

We comment that in all cases considered in this paper, the genera of
curves, if not zero, are always odd. We expect that this feature can
be understood by the loop-by-loop induction relation, in the future
work.

Unitarity cuts with fewer propagators are more complicated, since in
these cases, the on-shell equations define algebraic (hyper-)surfaces
instead of curves. 
The topological structures of surfaces are much more complicated than
those of curves. However, we expect that the techniques of
computational algebraic geometry would still be important for
obtaining the topological information of unitarity cuts. Analysis of
the global structure of generalized unitarity cuts is just the
beginning of multi-loop amplitude calculation,
and more works, such as parametrization or
the branch-by-branch polynomial fitting \cite{Badger:2012dv}, can be done based by this information.

\section*{Acknowledgment}
We thank Simon Badger, Simon Caron-Huot, Poul Damgaard, Hjalte
Frellesvig, Pierpaolo Mastrolia, David Skinner and Michael Stillman for useful discussion on this
project. Especially, we express gratitude to Simon Caron-Huot for
inspiring discussion on the three-loop genus computation and careful
reading of this paper in the draft stage. We also thank Pierpaolo
Mastrolia for his comments on this paper. RH wants to thank Bo Feng
and Center of Mathematical Science in Zhejiang University for the
hospitality during his visit. YZ is supported by Danish Council for
Independent Research-Natural Science (FNU) grant 11-107241.


\begin{thebibliography}{999}

\bibitem{Britto:2004ap}
  R.~Britto, F.~Cachazo and B.~Feng,
  Nucl.\ Phys.\ B {\bf 715}, 499 (2005)
  [hep-th/0412308].

\bibitem{Britto:2005fq}
  R.~Britto, F.~Cachazo, B.~Feng and E.~Witten,
  Phys.\ Rev.\ Lett.\  {\bf 94}, 181602 (2005)
  [hep-th/0501052].


\bibitem{Landau:1959fi}
  L.~D.~Landau,
  Nucl.\ Phys.\  {\bf 13}, 181 (1959);

  S.~Mandelstam,
  Phys.\ Rev.\  {\bf 112}, 1344 (1958);

  S.~Mandelstam,
  Phys.\ Rev.\  {\bf 115}, 1741 (1959);

  R.~E.~Cutkosky,
  J.\ Math.\ Phys.\  {\bf 1}, 429 (1960).

\bibitem{Bern:1994cg}
  Z.~Bern, L.~J.~Dixon, D.~C.~Dunbar and D.~A.~Kosower,
  Nucl.\ Phys.\  B {\bf 435}, 59 (1995)
  [arXiv:hep-ph/9409265].

\bibitem{Bern:1994zx}
  Z.~Bern, L.~J.~Dixon, D.~C.~Dunbar and D.~A.~Kosower,
  Nucl.\ Phys.\  B {\bf 425}, 217 (1994)
  [arXiv:hep-ph/9403226].

\bibitem{ArkaniHamed:2008gz}
  N.~Arkani-Hamed, F.~Cachazo and J.~Kaplan,
  JHEP {\bf 1009}, 016 (2010)
  [arXiv:0808.1446 [hep-th]].


\bibitem{Anastasiou:2006jv}
  C.~Anastasiou, R.~Britto, B.~Feng, Z.~Kunszt and P.~Mastrolia,
  Phys.\ Lett.\  B {\bf 645}, 213 (2007)
  [arXiv:hep-ph/0609191].

\bibitem{Anastasiou:2006gt}
  C.~Anastasiou, R.~Britto, B.~Feng, Z.~Kunszt and P.~Mastrolia,
  JHEP {\bf 0703}, 111 (2007)
  [arXiv:hep-ph/0612277].

\bibitem{Britto:2010xq}
  R.~Britto,
  J.\ Phys.\ A {\bf 44}, 454006 (2011)
  [arXiv:1012.4493 [hep-th]].


\bibitem{Britto:2004nc}
  R.~Britto, F.~Cachazo and B.~Feng,
  Nucl.\ Phys.\  B {\bf 725}, 275 (2005)
  [arXiv:hep-th/0412103].


\bibitem{Britto:2005ha}
  R.~Britto, E.~Buchbinder, F.~Cachazo and B.~Feng,
  Phys.\ Rev.\  D {\bf 72}, 065012 (2005)
  [arXiv:hep-ph/0503132].

\bibitem{IBP} F. V. Tkachov, Phys. Lett. B 100, 65 (1981); \\ K. G. Chetyrkin and
F. V. Tkachov, Nucl. Phys. B 192, 159 (1981). \\ S. Laporta, Phys.
Lett. B 504, 188 (2001) [hep-ph/0102032]. \\ S. Laporta, Int. J.
Mod. Phys. A 15, 5087 (2000) [hep-ph/0102033].


\bibitem{Ossola:2006us}
  G.~Ossola, C.~G.~Papadopoulos and R.~Pittau,
  Nucl.\ Phys.\ B {\bf 763}, 147 (2007)
  [hep-ph/0609007].



\bibitem{Forde:2007mi}
  D.~Forde,
  Phys.\ Rev.\ D {\bf 75}, 125019 (2007)  [arXiv:0704.1835 [hep-ph]].

\bibitem{Ellis:2007br}
  R.~K.~Ellis, W.~T.~Giele and Z.~Kunszt,
  JHEP {\bf 0803}, 003 (2008)  [arXiv:0708.2398 [hep-ph]].  


\bibitem{Kilgore:2007qr}
  W.~B.~Kilgore,
   arXiv:0711.5015 [hep-ph].  


\bibitem{Giele:2008ve}
  W.~T.~Giele, Z.~Kunszt and K.~Melnikov,
  JHEP {\bf 0804}, 049 (2008)  [arXiv:0801.2237 [hep-ph]].  


\bibitem{Ossola:2008xq}
  G.~Ossola, C.~G.~Papadopoulos and R.~Pittau,
  JHEP {\bf 0805}, 004 (2008)  [arXiv:0802.1876 [hep-ph]].  


\bibitem{Badger:2008cm}
  S.~D.~Badger,
  JHEP {\bf 0901}, 049 (2009)  [arXiv:0806.4600 [hep-ph]].  

\bibitem{Gluza:2010ws}
  J.~Gluza, K.~Kajda and D.~A.~Kosower,
  Phys.\ Rev.\ D {\bf 83}, 045012 (2011)
  [arXiv:1009.0472 [hep-th]].



\bibitem{Kosower:2011ty}
  D.~A.~Kosower and K.~J.~Larsen,
  Phys.\ Rev.\ D {\bf 85}, 045017 (2012)
  [arXiv:1108.1180 [hep-th]].

\bibitem{Larsen:2012sx}
  K.~J.~Larsen,
  arXiv:1205.0297 [hep-th].

\bibitem{CaronHuot:2012ab}
  S.~Caron-Huot and K.~J.~Larsen,
  arXiv:1205.0801 [hep-ph].  

\bibitem{Kleiss:2012yv}
  R.~H.~P.~Kleiss, I.~Malamos, C.~G.~Papadopoulos and R.~Verheyen,
   arXiv:1206.4180 [hep-ph].  


\bibitem{Johansson:2012zv}
  H.~Johansson, D.~A.~Kosower and K.~J.~Larsen,
  arXiv:1208.1754 [hep-th].

\bibitem{Johansson:2012sf}
  H.~Johansson, D.~A.~Kosower and K.~J.~Larsen,
  arXiv:1212.2132 [hep-th].



\bibitem{Mastrolia:2011pr}
  P.~Mastrolia and G.~Ossola,
  JHEP {\bf 1111}, 014 (2011)
  [arXiv:1107.6041 [hep-ph]].

\bibitem{Badger:2012dp}
  S.~Badger, H.~Frellesvig and Y.~Zhang,
  JHEP {\bf 1204}, 055 (2012)
  [arXiv:1202.2019 [hep-ph]].



\bibitem{Zhang:2012ce}
  Y.~Zhang,
  JHEP {\bf 1209}, 042 (2012)
  [arXiv:1205.5707 [hep-ph]].

\bibitem{Mastrolia:2012an}
  P.~Mastrolia, E.~Mirabella, G.~Ossola and T.~Peraro,
  Phys.\ Lett.\ B {\bf 718}, 173 (2012)
  [arXiv:1205.7087 [hep-ph]].


\bibitem{Feng:2012bm}
  B.~Feng and R.~Huang,
  arXiv:1209.3747 [hep-ph].

\bibitem{Mastrolia:2012wf}
  P.~Mastrolia, E.~Mirabella, G.~Ossola and T.~Peraro,
  arXiv:1209.4319 [hep-ph].

\bibitem{Mastrolia:2012du}
  P.~Mastrolia, E.~Mirabella, G.~Ossola, T.~Peraro and H.~van Deurzen,
  arXiv:1209.5678 [hep-ph].



\bibitem{Badger:2012dv}
  S.~Badger, H.~Frellesvig and Y.~Zhang,
  JHEP {\bf 1208}, 065 (2012)
  [arXiv:1207.2976 [hep-ph]].




\bibitem{MR0463157}
 R.~Hartshorne, {\sl Algebraic geometry}. Springer-Verlag, New York,
 1977. Graduate Texts in Mathematics, No. 52.

\bibitem{MR2372337}
    D.~Perrin, {\sl Algebraic geometry}. Universitext. Springer-Verlag London
    Ltd., London, 2008. An introduction, Translated from the 1995 French original by Catriona
    Maclean.

\bibitem{M2}
D. R.~Grayson, M.E.~Stillman, "Macaulay2, a software system for
research in algebraic geometry." Available at
http://www.math.uiuc.edu/Macaulay2/.

\bibitem{MR2290010}
    D.~Cox, J. ~Little, and D.~O'Shea, {\sl Ideals, varieties, and
    algorithms}. Undergraduate Texts in Mathematics. Springer, New
    York, Third ed., 2007. An introduction to computational algebraic geometry and commutative algebra.


\end{thebibliography}
\end{document}